%% file: manuscript.tex
\documentclass{svjour3} % onecolumn (second format)
\smartqed  % flush right qed marks, e.g. at end of proof
\usepackage{natbib}
\usepackage{tikz}
\usepackage{graphicx}
\usepackage{lmodern}
\usepackage{csquotes}
\usepackage{microtype}
\usepackage{multirow}
\usepackage{amsmath}
\usepackage{booktabs}
\usepackage{amsfonts}
\usepackage{bigstrut}
\usepackage{color}
\newcommand{\subparagraph}{}
\usepackage{titlesec}
\usepackage{float}
\usepackage{graphics}
\usepackage{colortbl}
\usepackage[scaled=1]{helvet} 
\usepackage{balance}
\usepackage[shortlabels]{enumitem}
\usepackage{url}
\usepackage{rotating}
\usepackage[framed]{ntheorem}
\usepackage{framed}
\usepackage[hidelinks]{hyperref}
\usepackage{subfig}
\usepackage{soul}
\usepackage[font=small]{caption}
\usepackage{algorithm}
\usepackage{algpseudocode}
\usepackage{amssymb}
\makeatletter
\algrenewcommand\ALG@beginalgorithmic{\footnotesize}
\makeatother

\definecolor{lightgray}{gray}{0.8}
\definecolor{darkgray}{gray}{0.5}
\definecolor{lightgreen}{rgb}{.68, .79, .46}
\definecolor{gr1}{HTML}{F0F0F0}
\definecolor{gr2}{HTML}{DDDDDD}
\definecolor{gr3}{HTML}{CCCCCC}
\definecolor{gr4}{HTML}{BBBBBB}

\definecolor{lavenderpink}{HTML}{F3F3F3}
\definecolor{celadon}{gray}{0.7}
\definecolor{Gray}{gray}{0.85}
\definecolor{LightGray}{gray}{0.975}
\definecolor{cellgrey}{rgb}{.87, .87, .87}
\definecolor{steel}{RGB}{65, 105, 225}
\definecolor{shadecolor}{gray}{0.95}

\newcommand{\result}[1]{
\noindent
\begin{tabular}{p{0.95\linewidth}}
\cellcolor{Gray}\textit{\textbf{\underline{Result}:}}~#1
\end{tabular}
}
\newcommand{\bi}{\begin{itemize}} %[leftmargin=0.4cm]}
\newcommand{\ei}{\end{itemize}}
\newcommand{\be}{\begin{enumerate}}
\newcommand{\ee}{\end{enumerate}}
\newcommand{\ktest}{$\mathbb{K}$-test}
\newcommand{\tion}[1]{\S~\ref{sect:#1}}
\newcommand{\fig}[1]{Fig.~\ref{fig:#1}}
\newcommand{\tab}[1]{Table ~\ref{tab:#1}}
\newcommand{\eq}[1]{Equation~\ref{eq:#1}}
\renewcommand{\tt}[1]{\texttt{#1}}

\newcommand{\compactlist}{
 \begin{list}{$\bullet$}
 { \setlength{\itemsep}{0pt}
  \setlength{\parsep}{0pt}
  \setlength{\topsep}{0pt}
  \setlength{\partopsep}{0pt}
  \setlength{\leftmargin}{0em}
  \setlength{\labelwidth}{1.5em}
  \setlength{\labelsep}{0.5em} } }

\newcommand{\compactend}{
\end{list}}

\titlespacing*{\section}{0pt}{0.8\baselineskip}{0.8\baselineskip}
\titlespacing*{\subsection}{0pt}{0.7\baselineskip}{0.7\baselineskip}
\titlespacing*{\subsubsection}{0pt}{0.66\baselineskip}{0.66\baselineskip}
\setitemize{noitemsep,topsep=0pt,parsep=0pt,partopsep=0pt,wide=0pt,leftmargin=*}
\setenumerate{noitemsep,topsep=0pt,parsep=0pt,partopsep=0pt,wide=0pt,leftmargin=*}
\begin{document}
\title{Learning Actionable Analytics from \\Multiple Software Projects}
\author{Rahul Krishna \and Tim Menzies}

\institute{Rahul Krishna \at
Computer Science \\
NC State University \\
\email{i.m.ralk@gmail.com}%
\and
Tim Menzies \at
Computer Science \\
NC State University \\
\email{timm@ieee.org}%
}

\date{Received: date / Accepted: date}
\maketitle

\input{0_abstract.tex}
\input{1_introduction.tex}
\input{2_rqs.tex}

\input{3_motivation.tex}

\input{4_planning.tex}

\input{5_xtree.tex}

\input{6_methods.tex}

\input{7_results.tex}

\input{8_discuss.tex}

\input{9_threats.tex}

\input{10_conclusion.tex}

\section*{Acknowledgements}
The work is partially funded by NSF awards \#1506586 and \#1302169.

% --------------------------
% ------ References --------
% --------------------------
\bibliographystyle{spbasic}
\bibliography{references}
\end{document}

%% file: 0_abstract.tex
\begin{abstract}

    The current generation of software analytics tools are mostly prediction algorithms (e.g. support vector machines, naive
    bayes, logistic regression, etc). While prediction is useful, after prediction comes {\em planning} about what actions to take in order to
    improve quality. This research seeks methods that generate demonstrably useful guidance on ``what to do'' within the context of a
    specific software project. Specifically, we propose XTREE (for within-project planning) and BELLTREE (for cross-project planning) to
    generating plans that can improve software quality. Each such plan has the property that, if followed, it reduces the expected number of future
    defect reports.
    To find this expected number,   planning was first applied to data from release $x$. Next,  we looked for change  in release $x+1$
    that conformed to our plans. 
    This procedure was applied using a range of planners from the literature, as well as  XTREE. 
    In 10 open-source JAVA systems, several hundreds of defects
    were reduced in sections of the code that conformed to XTREE's  plans.
    Further, when compared to other planners,
    XTREE's plans were found to be easier to implement (since they were shorter) and more effective at reducing the expected number of defects.
    % across projects, which is particularly useful when there are no historical logs available for a particular project to generate plans from.
    \keywords{Data Mining, Actionable Analytics, Planning, bellwethers, defect prediction.}
    \end{abstract}
    

%% file: 1_introduction.tex
\newpage
\section{Introduction}
Data mining tools have been succesfully applied to many applications in software engineering; e.g.~\citep{czer11, ostrand04, Menzies2007a, turhan11, koc11b, export:208800, theisen15}. 
Despite these successes,  current
software analytic tools have certain drawbacks. At a workshop on ``Actionable Analytics'' at the 2015 IEEE conference on
Automated Software Engineering, 
business users were vocal in their complaints about analytics~\citep{hihn15}.
``Those tools tell us \textit{what is}, '' said one business user, ``But they don't tell us \textit{what to do}''.
Hence we seek new tools that offer  guidance on ``what to do'' within a specific project. 

We seek such new tools since  current   analytics tools are mostly \textit{prediction} algorithms such as support vector machines~\citep{cortes95}, naive Bayes classifiers~\citep{lessmann08}, logistic regression~\citep{lessmann08}. For example, defect prediction tools report what combinations of software project features predict for some dependent variable (such as the number of defects). Note that this is a different task to \textit{planning}, which answers the question: what to {\em change} in order to {\em improve} quality.
	
More specifically, we seek plans that propose {\em least} changes while most \textit{improving} software \textit{quality} where:
\bi
\item \textit{Quality} = defects reported by the development team; 
\item \textit{Improvement} = lowered likelihood of future defects.
\ei
This paper advocates the use of the {\em bellwether effect}~\citep{krishna16, krishna17a, mensah2018investigating} to generate plans. This effect states that:
\begin{quote}
  \textit{`` \ldots When a community of programmers work on a set of projects, then within that community there exists one exemplary project, called the bellwether\footnote{According to the Oxford English Dictionary, the bellwether is the leading sheep of a flock, with a bell on its neck.}, which can best define quality predictors for the other projects \ldots ''}
\end{quote}
Utilizing the bellwether effect, we propose a cross-project variant of our XTREE contrast set learner called BELLTREE where
\begin{center}
  \textit{BELLTREE} = \textit{Bellwether}$+$\textit{XTREE}
\end{center}
BELLTREE searches for an exemplar project, or \textit{bellwether}~\citep{krishna17a}, to construct plans from other projects. As shown by
the experiments of this paper, these plans can be remarkably effective. In 10 open-source JAVA systems, hundreds of defects could potentially be reduced in sections of the code that followed the plans generated by our planners. Further, we show that planning is possible across projects, which is particularly useful when there are no historical logs available for a particular project to generate plans from.

The structure of this paper is as follows: the rest of this section highlights the key contributions of this work(\tion{contrib}), and relationships between this work and our prior work (\tion{our_prior}). In \tion{rqs}, we introduce the research questions asked in this paper and briefly discuss our findings. In \tion{motivate} we discuss the background which include some of related work in the area. There, in \tion{planners}, the  notion of planning and the different kinds of planners studied here. \tion{prelim} contains the research methods, datasets, and evaluation strategy. In \tion{results} we answer the research questions. In \tion{discuss} we discuss the implications of our findings. Finally, \tion{threats} and \tion{future} present threats to validity and conclusions respectively.

\subsection{Contributions}
\label{sect:contrib}
The key contributions of this work are:

\textit{1. New kinds of software analytics techniques:} This work combines planning~\citep{krishna17a} with cross-project learning using bellwethers~\citep{krishna16}. 
Note that
our previous work ~\citep{krishna16, krishna17b} explored prediction and not the planning as described here. Also, previously, our planners~\citep{krishna17a} only explored within-project problems (but not cross-project). 

\textit{2. Compelling results about planning:} Our results show that planning is  successful in producing actions that can reduce the number of defects; Further, we see that plans learned on one project can be translated to other projects.

\textit{3. More evidence of generality of bellwethers:}  Bellwethers were
originally  used in the context of prediction~\citep{krishna16} and have been shown to work for (i)~defect prediction, (ii)~effort estimation, (iii)~issues close time, and (iv)~detecting code smells~\citep{krishna17b}. This paper extends those results to show that bellwethers can also be used from cross-project planning. This is an important result of much significance since, it suggests that general conclusions about SE can be easily found (with bellwethers).

\textit{4. An open source reproduction package containing all our scripts and data.} For readers interested in replicating this work, kindly see \url{https://git.io/fNcYY}.

\subsection{Post Hoc Ergo Propter Hoc?}\label{sect:hoc}

The Latin expression {\em post hoc ergo propter hoc}
 translates to ``after this, therefore because of this''.
This Latin expression is the name given to
the logical fallacy that
``since event Y followed event X, event Y must have been caused by event X''.
This can be a fallacy since another event Z may have influenced Y.

This concern was very
present in our minds as we developed this paper.
Prior to this paper, it was an open issue if 
XTREE/BELLTREE's plans  work on future data. Accordingly
we carefully evaluated if knowledge of past changes were  useful for 
planning future changes.
The details of that evaluation criteria are offered later in this paper
(see ``The {\ktest}'' of \S\ref{sect:ktest}).  For now, all we need say is that:
\bi
\item
We sorted our data via its associated timestamps
into {\em older}, {\em newer}, and {\em latest}
(later in this paper we will call these {\em train, test, validate}, respectively).
We say that the 
{\em older} plans are those learned from   the  {\em older data}. 
\item
If developers of   {\em newer} code knew about  the  older plans,
then they would  apply them
either (a)~{\em very little}, (b)~{\em some}, (c)~{\em more}; or (d)~{\em mostly}. 
\item
We also note that  it is possible to  automatically identify
each of those four kinds developers  as those whose changes  
between {\em newer} and {\em latest}
overlap
with  the older plans  (a)~{\em very little}, (b)~{\em some}, (c)~{\em more}; or (d)~{\em mostly}. 
\ei
The experiments of this paper show that,
 when we explored real world data from
from the {\em newer} and {\em latest} versions, then:
\bi
\item If projects changes overlap  {\em very little } 
with  older plans, then   defects are not reduced.
\item But if projects changes {\em mostly} overlap   with   older plans, then   defect  are  much lower.
\ei
To be clear,   XTREE/BELLTREE {\em does not} generate causal
models for software defects. However, our results  suggest that it can be very useful to follow
our plans.

\subsection{Relationship to Prior Work}
\label{sect:our_prior}
As for the connections to prior research, 
as shown in \fig{analytics}, originally in 2007 we explored software quality prediction in the context of training and testing within the same software project~\citep{menzies07}. After that we found ways in 2009 to train these predictors on some projects, then test them on others~\citep{turhan09}. Subsequent work in 2016 found that bellwethers were a simpler and effective way to implement transfer learning~\citep{krishna16}, which worked well for a wide range of software analytics tasks~\citep{krishna17b}. 

\input{contributions}
In the area of planning, we introduced the possibility of using XTREE for planning as a short report at a workshop on \enquote{Actionable Analytics} in ASE `15~\citep{krishna15}, we followed this up a slightly more detailed report in the IST journal~\citep{krishna2017less}. These initial findings on XTREE were also presented at the IEEE ASE'17 Doctoral Symposium~\citep{krishna2017b}. The panel highlighted the following limitations:
\bi[wide=0pt]
\item \textit{Inadequate Validation.} Our initial report uses \textit{defect predictors} to assess    
plan effectiveness. However, the  performance of those defect prediction schemes were limited to at most 65\% (as shown in Figure 5 of~\citep{krishna2017less}).   
\item \textit{Smaller Datasets.} Due to the limited predictive performance of the defect predictors used in the previous studies, the results were reported on only five projects.
\item \textit{Metric interdepencies ignored.} The previous variant of XTREE also did not take into consideration the interaction between individual metrics. 
\ei
Accordingly, in this paper we present a updated variant of XTREE, including new experiments on more projects.

Further, this current article     addresses a much harder question: can plans be generated from one project and applied to the another? In answering this, we have endeavored to avoid our mistakes from the past, e.g., the use of overly complex methodologies to achieve a relatively simpler goal. Accordingly, this work experiments with bellwethers to see if this simple method works for planning as with prediction. 

One assumption across much of our work is the \textit{homogeneity} of the learning, i.e., although the training and testing data may belong to different projects, they share the same attributes~\citep{krishna16, krishna17a, krishna17b, menzies07, turhan09}. Since that is not always the case, we have recently been exploring heterogeneous learning where attribute names may change between the training and test sets~\citep{fu18}. Heterogeneous planning is primary focus of our future work.
% \newpage

%% file: contributions.tex
\begin{figure}[!b]
\centering
\resizebox{\linewidth}{!}{
\begin{tabular}{lccc}

  \cline{2-4}

\multicolumn{1}{l|}{} & \multicolumn{1}{c|}{Data source= Within} & \multicolumn{2}{c|}{Data source = Cross} \bigstrut\\ \cline{1-4}

\multicolumn{1}{l|}{\multirow{3}{*}{Prediction}} & \multicolumn{1}{c|}{\multirow{3}{*}{TSE '07~\citep{menzies07}}} & \multicolumn{1}{l|}{EMSE '09~\citep{turhan09}}  & \multicolumn{1}{c|}{\multirow{3}{*}{TSE '17~\citep{fu18}}}   \bigstrut\\

\multicolumn{1}{l|}{}                            & \multicolumn{1}{c|}{}                         & \multicolumn{1}{l|}{ASE '16~\citep{krishna16}}   & \multicolumn{1}{c|}{}                           \bigstrut\\

\multicolumn{1}{l|}{}                            & \multicolumn{1}{c|}{}                         & \multicolumn{1}{l|}{TSE '18~\citep{krishna17b}}   & \multicolumn{1}{c|}{}                           \bigstrut\\ \cline{1-4}

\multicolumn{1}{l|}{Planning}                    & \multicolumn{1}{c|}{IST '17~\citep{krishna17a}}                  &  { \cellcolor{lightgray}This work}    &
\multicolumn{1}{l|}{Future work}   \bigstrut\\ \hline

\multicolumn{1}{l|}{}                            & \multicolumn{2}{c|}{Homogeneous}                                               & \multicolumn{1}{c|}{Heterogeneous}              \bigstrut\\ \cline{2-4}

\end{tabular}}
 \caption{Relationship of this paper to our prior research. 
%Within project trained and tested data miners using data from the same project. Cross projects train on one project, then test on another. Homogeneous learning requires the attribute names to be identical in the training and test set. Heterogeneous learning relaxes that requirement; i.e. the attribute names might change from the training to the test set.
}
\label{fig:analytics}
 \end{figure}

%% file: 2_rqs.tex
\section{Research Questions}
\label{sect:rqs}
The work in this paper is gudeied by the following research questions.

\subsubsection*{RQ1: How well do planners' recommendations match developer actions?}
\compactlist
    \item[]\textit{\underline{Motivation}:}~There is no point offering plans that no one will follow.
    Accordingly, on this research question, we ask how many of a planner's recommendations match with the actions taken by developers to fix defects in their files.

    \item[]\textit{\underline{Approach}:}~We measure the \textit{overlap} between the planners' recommendations developers' actions. Then, plot the aggregate number files for overlap values ranging from 0\% to 100\% in bins of size 25\% (for ranges of $0-25\%$, $26-50\%$, $51-75\%$, and $76-100\%$). Planners that have the larger aggregate number files for higher overlap ranges are considered better.

    \item[]\textit{\underline{Evaluation}:}~We compare XTREE with three other outlier statistics based planners from current literature namely, those of \cite{alves}, \cite{shatnawi}, and \cite{oliveira}.
    % \smallskip
    \item[]\result{XTREE significantly outperforms all other outlier statistics based planners. Further, in all the projects studied here, most of the developers actions to fix defects in a file has a $76-100\%$ overlap with the recommendations offered by XTREE.}
    \compactend
    
\subsubsection*{RQ2: Do planners' recommendations lead to reduction in defects?}
\compactlist
    \item[]\textit{\underline{Motivation}:}~The previous research question measured the extent to which a planner's recommendations matched the actions taken by developers to fix defects in their files. But, a high overlap in most files does not necessarily mean that the defects are actually reduced. Likewise, it is also possible that defects are added due to other actions the developer took during the development. Thus, here we ask how many defects are reduced, and how many are added, in response to larger overlap with the planners' recommendations.

    \item[]\textit{\underline{Approach}:}~Like before, we measure the \textit{overlap} between the planners' recommendations developers' actions. Then, we plot the aggregate number defects reduced in file with overlap values ranging from 0\% to 100\% in bins of size 25\% (for ranges of $0-25\%$, $26-50\%$, $51-75\%$, and $76-100\%$). Planners that have a large number defects reduced for higher overlap ranges are considered better.
 
    \item[]\textit{\underline{Evaluation}:}~Similar to RQ1, we compare XTREE with three other outlier statistics based planners of Alves et al., Shatnawi, and Oliveira, for the overall number of defects reduced and number of defects added.
    % \smallskip
    \vspace{0.2em} 

    \item[]\result{Plans generated by XTREE are superior to other outlier statistics based planners in all 10 projects. Planning with XTREE leads to the far larger number of defects reduced as opposed to defects added in 9 out of 10 projects studied here.}
\compactend

\subsubsection*{RQ3: Are cross-project plans generated by BELLTREE as effective as within-project plans of XTREE?}
\compactlist
    \item[]\textit{\underline{Motivation}:}~The previous research questions we assume that there exists historical data to construct the planning algorithms. However, given the pace of software change, for new projects, it is quite possible that there is insufficient historical data to perform planning. Thus, this research question asks if it is possible to use data from other software projects to construct planners to generate recommendations.
    \item[]\textit{\underline{Approach}:}~We use a cross-project planner that discovers the \textit{bellwether} dataset. Using this bellwether project, we construct XTREE as generate plans as usual. We refer to this combination of using Bellwethers with XTREE as BELLTREE.
    \item[]\textit{\underline{Evaluation}:}~Here we compare BELLTREE with a conventional XTREE and with one other outlier statistics based planner (Shatnawi) to measure the number of defects reduced and number of defects added.
    \vspace{0.2em} 
    % \smallskip
    \item[]\result{The effectiveness of BELLTREE is comparable to the effectiveness of XTREE. In 8 out of 17 BELLTREE outperformed XTREE and 9 out of 17 cases, XTREE outperformed BELLTREE. BELLTREE and XTREE outperformed other planners in all cases.}
\compactend

%% file: 3_motivation.tex
\section{Motivation}
\label{sect:motivate}
\subsection{Defect Prediction}
\label{sect:defect_prediction}

As projects evolve with additional functionalities, they also add defects, as a result the software may crash (perhaps at the most inopportune time) or  deliver incorrect or incomplete functionalities. Consequently, programs are tested before deployment. However, an exhaustive testing is expensive and software assessment budgets are finite~\citep{LowryBK98}.  Exponential  costs  quickly  exhaust  finite resources, so  standard  practice  is  to  apply  the  best  available methods only on code sections that seem most critical. 

One approach is to use defect predictors learned from static code metrics. Given software described in terms of the metrics of~\tab{static_metrics}, data miners can learn where the probability of software defects is the highest. These static code metrics can be automatically collected, even for very large systems~\citep{nagappan05}. Further, these static code metrics based defect predictors can be quickly adapted to new languages by building lightweight parsers to computes metrics similar to that of~\tab{static_metrics}. Over the past decade, defect predictors have granered a significant amount of interest. They are frequently reported to be capable of finding the locations of over  70\% (or more) of the defects in code~\citep{me07b,Nam13,fu,ghotra2015revisiting,lessmann08,fu18,krishna17b}. Further, these defect predictors seem to have some level of generality~\cite{Nam13,Nam15,krishna16,krishna17b}. The success of these methods in finding bugs is markedly higher than other currently-used industrial methods such as manual code reviews~\citep{shu02}. Although other methods like  manual code reviews are much more accurate in identifying defects, they take much higher effort to find a defect and also are relatively slower. For example, depending on the review methods, 8 to 20 LOC/minute can be inspected and this effort repeats for all members of the review team, which can be as large as four or six people~\citep{me02f}. For these reasons, researchers and industrial practitioners use static code metrics to guide software  quality predictions. Defect prediction has been favored by organizations such as Google~\cite{lewis13} and Microsoft~\citep{Zimmermann09}. 

  Although the ability to predict defects in software systems is viewed favorably by researchers and industrial practitioners, the current generation of defect prediction is subject to several criticisms. There is are open debates on the efficacy of static code metrics and the existence of causal links between these metrics and the defect counts. While a number of studies favor static code metrics, there are some that prefer other type of metrics. We explore these in greater detail in  \tion{metrics}.

Another major criticism of software defect prediction is that they lack actionable guidance, i.e., while these techniques enable developers to target defect-prone areas faster, but do not guide developers toward a particular action that leads to a fix. Without a such guidance, developers are often tasked with making a majority of the decisions. However, this could be problematic since researchers have cautioned that developers' cognitive biases often leads to misleading assertions on how best to make a change. For instance, Passos et al.~\citep{passos11} remarks that developers often assume that the lessons they learn from a few past projects are general to all their future projects.  They comment, ``past experiences were taken into account without much consideration for their context''~\citep{passos11}. Such warnings are also echoed by J{\o}rgensen \& Gruschke~\citep{jorgensen09}. They report that the supposed software engineering experts seldom use lessons from past projects to improve their future reasoning and that such poor past advice can be detrimental to new projects. Other studies have shown that some widely-held views are now questionable given new evidence. Devanbu et al. observes that, on examination of responses from 564 Microsoft software developers from around the world, the programmer beliefs can vary significantly with each project, but that these beliefs do not necessarily correspond with actual evidence in that project~\citep{prem16}.

For the above reasons, in this paper, we seek newer analytics tools that go beyond traditional defect prediction to offer ``plans''. Instead of just pointing to the likelihood of defects, these ``plans'' offer a set of changes that can be implemented to reduce the likelihood of future defects. We explore the notion of planning in greater detail in the following section (see~\tion{planning_intro}). 

\subsection{Choice of Software Metrics}
\label{sect:metrics}
\input{metrics.tex}

The data used in our studies use \textit{static code metrics} to quantify the aspects of software design. These metrics have been measured in conjunction with faults that are recorded at a number of stages of software development such as during requirements, design, development, in various testing phases of the software project, or with a post-release bug tracking systems. Over the past several decades, a number of \textit{metrics} have been proposed by researchers for the use in software defect prediction. These metrics can be classified into two  categories: (a) Product Metrics, and (b) Process Metrics.

Product metrics are a syntactic measure of source code in a specific snapshot of a software project. The metrics consist of McCabe and Halstead complexity metrics, LOC (Lines of Code), and Chidamber and Kemerer Object-Oriented (CK OO) metrics as shown in as shown in~\tab{static_metrics}.~\cite{mccabe1976complexity} and~\cite{halstead77} metrics are a set of static code metrics that provide a quantitative measure of the code complexity based on the decision structure of a program. The idea behind these metrics is that the more structurally complex a code gets, the more difficult it becomes to test and maintain the code and hence the likelihood of defects increases. McCabe and Halstead metrics are well suited for traditional software engineering and are inadequate in and of themselves. To measure aspects of object oriented (OO) design such as classes, inheritance, encapsulation, message passing, and other unique aspects of OO approach,~\citep{chidamber1994metrics} developed as set of OO metrics. When used in conjunction with McCabe and Halstead metrics, these measures lend themselves to a more comprehensive analysis.

Process metrics differ from product metrics in that they are computed using the data obtained from change and defect history of the program. Process metrics measure such aspects as the number of commits made to a file, the number of developers who changed the file, the number of contributors who authored less than 5\% of the code in that file, the experience of the highest contributor. All these metrics attempt to comment on the software development practice rather than the source code itself. 

The choice of metrics from the perspective of defect prediction as has been a matter of much debate.  In recent years, a number of researchers and industrial practitioners (at companies such as Microsoft) have demonstrated the effectiveness of  static code metrics to build predictive analytics. A commonly reported effect by a number of researchers like~\citep{al2010object,shatnawi2008effectiveness,madeyski2015process,chidamber1998managerial,menzies07,alves,bener2015lessons,shatnawi,oliveira} is that OO metrics show a strong correlation with fault proneness. A comprehensive list of research on the correlation between product metrics and fault proneness can be found in Table 1 of the survey by~\citep{Rathore2019}.

Some researchers have criticized the use of static code metrics to learn defect predictors. For instance,~\citep{graves2000} critiqued their effectiveness due to the fact that many metrics are highly correlated with each other, while~\citep{rahman2013} claim that static code metrics may not evolve with the changing distribution of defects, which leads code-metric-based prediction models becoming stagnated. However, on close inspection of both these studies, we noted that some of the most informative static code metrics have not been accounted for. For example, in the case of~\citep{graves2000}, they only inspect the McCabe and Halstead metrics and not object oriented metrics. In the case of~\citep{rahman2013}, (a) 37 out of 54 static code metrics (over $\frac{2}{3}$) are file-level metrics, most of which are not related to OO design, and (b) many of the metrics are repeated variants of the same measure (e.g., $CountLineCode$, $RatioCommentToCode$, $CountLineBlank$, etc are all measure of lines of code in various forms).

Given this evidence that static code metrics relate to defects, we use these metrics for our study. The defect dataset used in the rest of this this paper comprises a total of 38 datasets from 10 different projects taken from previous transfer learning studies. This group of data was gathered by Jureczko et al. \citep{Jureczko2010}. They recorded the number of known defects for each class using a post-release bug tracking system. The classes are described in terms of 20 OO metrics, including extended CK metrics, McCabes and complexity metrics, see~\tab{static_metrics} for description. We obtained the dataset from the SEACRAFT repository\footnote{$\text{https://zenodo.org/communities/seacraft/}$} (formerly the PROMISE repository~\citep{menzies2016promise}). 

\input{datasets.tex}

%% file: metrics.tex
\begin{table*}[!t]
    \caption{Sample static code attributes.}\label{tab:static_metrics}
	\arrayrulecolor{black}
	\centering
	\resizebox{\linewidth}{!}{%
		\begin{tabular}{ll}
			\toprule
			\textbf{Metric} & \textbf{Description}  \\\midrule
			
			wmc & weighted methods per class  \\
			
			dit & depth of inheritance tree  \\
			
			noc &  number of children  \\
			
			cbo & increased when the methods of one
			class access services of another. \\
			
			rfc & number of  methods invoked in response to
			a message to the object. \\
			
			lcom &number of pairs of methods that do
			not share a reference to an instance variable. \\
			ca & how many other classes use the specific
			class.  \\
			
			ce & how many other classes is used by the
			specific class.  \\
			
			npm & number of public methods   \\
			
			locm3 & if $m,a$ are  the number of
			$methods,attributes$
			in a class number and $\mu(a)$  is the 
			\\
		    & number of methods accessing an
			attribute, 
			then
			$lcom3=((\frac{1}{a} \sum_j^a \mu(a_j)) - m)/ (1-m)$.
			\\
			
			loc & lines of code  \\
			
			dam & ratio of  private (protected)
			attributes to   total   attributes \\
			
			moa &  count of the number of data declarations (class
			fields) whose types are user defined classes \\
			
			mfa & number of methods inherited by a class
			plus number of \\
			& methods accessible by member methods of the
			class \\
			
			cam & summation of number of different
			types of method parameters in every method \\
			& divided by a multiplication
			of number of different method parameter types  \\ 
			& in whole class and
			number of methods.  \\
			
			ic &  number of parent classes to which a given
			class is coupled (includes counts \\
			& of methods and variables inherited)
			\\
			
			cbm &  total number of new/redefined methods
			to which all the inherited methods are coupled \\
			a
			mc & average methods oer class \\
			max\_cc & maximum McCabe's cyclomatic complexity seen
			in class \\
			
			avg\_cc & average McCabe's cyclomatic complexity seen
			in class \\\midrule
			defect & Defects found in post-release bug-tracking systems.\\\toprule
		\end{tabular}
	} 
\end{table*}

% \begin{figure}
% 	\renewcommand{\baselinestretch}{0.7}\begin{center}
% 		\resizebox{0.8\linewidth}{!}{
% 			\begin{tabular}{c|l}
% 				amc & average method complexity \\
% 				avg\, cc & average McCabe \\
% 				ca & afferent couplings \\
% 				cam & cohesion amongst classes \\
% 				cbm & coupling between methods \\
% 				cbo & coupling between objects \\
% 				ce & efferent couplings \\
% 				dam & data access\\
% 				dit & depth of inheritance tree\\
% 				ic & inheritance coupling\\
% 				lcom (lcom3) & 2 measures of lack of cohesion in methods \\
% 				loc & lines of code \\
% 				max\, cc & maximum McCabe\\
% 				mfa & functional abstraction\\
% 				moa &  aggregation\\
% 				noc &  number of children\\
% 				npm & number of public methods\\
% 				rfc & response for a class\\
% 				wmc & weighted methods per class\\
% 				\rowcolor{lightgray}
% 				\#defects & raw defect counts\\
% 			\end{tabular}
% 		}
% 	\end{center}
% 	\caption{OO code metrics used for all studies in this paper.
% 	   Last line, shown in \colorbox{lightgray}{gray}, denotes the dependent variable. For more details, see~\cite{krishna17a}.}\label{fig:static_metrics}
% \end{figure}

%% file: datasets.tex
\newcolumntype{M}{@{}>{\columncolor{white}[0pt][0pt]}l@{}}
\begin{figure}[!t]
\centering
\begin{minipage}{\linewidth}
\resizebox{\linewidth}{!}{%
\scriptsize \begin{tabular}{Mrrrl}
\toprule
Dataset & \multicolumn{1}{r}{Versions} & \multicolumn{1}{r}{$N$} & Bugs (\%) & Description\bigstrut\\\midrule
 Lucene & 2.2 -- 2.4 & 782  & 438 (56.01) & Information retrieval software library\bigstrut\\
 Ant   & 1.3 -- 1.7 & 1692  & 350 (20.69) & A software tool for automating\bigstrut\\
 &&&& software build processes\bigstrut\\
 Ivy   & \multicolumn{1}{r}{1.1, 1.4,2.0} & 704   & 119 (16.90) & A transitive package manager\bigstrut\\
 Jedit & \multicolumn{1}{r}{4.0 -- 4.3} & 1749  & 303 (17.32) & A free software text editor\bigstrut\\
 Poi   & \multicolumn{1}{r}{1.5, 2, 2.5, 3.0} & 1378  & 707 (51.31) & Java libraries for manipulating files in\bigstrut\\
 &&&& MS Office format.\bigstrut\\
 Camel & \multicolumn{1}{r}{1.0, 1.2, 1.4,1.6} & 2784  & 562 (20.19) & A framework for message-oriented middleware.\bigstrut\\
 Log4j & \multicolumn{1}{r}{1.0, 1.1,1.2} & 449   & 260 (57.91) & A Java-based logging utility.\bigstrut\\
 Velocity & \multicolumn{1}{r}{1.4, 1.5,1.6} & 639   & 367 (57.43) & A template engine to reference objects in Java.\bigstrut\\
 Xalan & \multicolumn{1}{r}{2.4, 2.5, 2.6,2.7} & 3320  & 1806 (54.40) & A Java implementation of XLST, XML, and XPath.\bigstrut\\
 Xerces & \multicolumn{1}{r}{1.0, 1.2, 1.3,1.4} & 1643  & 654 (39.81) & Software libraries for manipulating XML.\bigstrut\\\bottomrule
\end{tabular}%
}
\end{minipage}
\caption{The figure lists defect datasets used in this paper.}
\label{fig:datasets}
\end{figure}

%% file: 4_planning.tex
\section{What is Planning?}
\label{sect:planning_intro}
We distinguish planning from prediction for software quality as follows: 
Quality prediction points to the likelihood of defects. Predictors take the form:
\begin{equation*}
  out = f(in)  
\end{equation*}
where {\em in} contains many independent features (such as OO metrics) and {\em out} contains some measure of
how many defects are present. For software analytics, the function $f$ is learned via mining static code attributes.

\input{example.tex}

On the other hand, quality planning seeks  precautionary measures to significantly reduce the likelihood of future defects.

For a formal definition of plans, consider a defective test example $Z$, a planner
proposes a plan ``$\Delta$'' to adjust attribute $Z_j$ as follows:

{\small\[
\forall \delta_j \in \Delta : Z_j = 
\begin{cases}
   Z_j \pm \delta_j& \text{if $Z_j$ is numeric}\\
  \delta_j       & \text{otherwise}
\end{cases}
\]}

The above plans are described in terms of a range of numeric values. In this case, they represent an increase (or decrease) in some of the static code metrics of \tab{static_metrics}. However, these numeric ranges in and of themselves may not very informative. It would be beneficial to offer a more detailed report on how to go about implementing these plans. For example, to (say) simplify a large bug-prone method, it may be useful to suggest to a developer to reduce its size (e.g., by splitting it across two simpler functions).

In order to operationalize such plans, developers need some guidance on what to change in order to achieve the desired effect. 
There are two places to look
for that guidance:
\be
\item In other projects;
\item In the current project.
\ee
As to  the first approach
({\em using other projects}), 
 several recent papers have discussed how code changes adjust
static code metrics~\citep{stroggylos2007, du2006study, kataoka2002, bryton2009, elish2011, elish2012}.
For example,
\fig{motivating_example}(b) shows a summary of
that research. 
We could apply those results
as follows:\bi
\item
Suppose a planner has recommended the changes shown in \fig{motivating_example}(a). 
\item
Then, we use \ref{fig:motivating_example}\protect\subref{subfig:actions} to look-up possible actions developers may take.
Here, we see that performing an ``extract method'' operation may help alleviate certain defects (this is highlighted in {\colorbox{lightgray}{gray}}).
\item
In \ref{fig:motivating_example}\protect\subref{subfig:before} we show a simple example of a class where the above operation may be performed. 
\item
In \ref{fig:motivating_example}\protect\subref{subfig:after}, we demonstrate how a developer may perform the ``extract method''. 
\ei
While using other projects may be useful,
that approach has a problem. Specifically: 
what happens
 if the proposed change has not been studied before in the literature?
 For this reason,
 we prefer to use the second approach
 (i.e. {\em use the current project}).
 In that approach, we look through the developer's own history to find old examples where they have made the kinds of changes recommended by the plan. 
Other researchers also adopt this approach (see
\citep{nayrolles2018clever} at MSR 2018). In the following:
 \bi
 \item
Using frequent itemset  mining, we
   summarize prior  changes in the current
 project (for details on this kind of learning, see 
  \fig{xtree}.C).
 \item
Next, when we learn plans, we reject any
that are not known prior changes.
\ei
 In this way, we can ensure that if a developer asks ``how do I implement this plan?'',
 we can reply with a relevant
 example of prior changes to the current project.

\subsection{Planning in Software Engineering}
\label{sect:planners}

We say that \fig{motivating_example} is an example of {\em code-based planning} where the goal is to change a code base in order to improve that code in some way. The rest of this section first discusses other kinds of planning before discussing \textit{code based planning} in greater detail.

Planning is extensively explored in artificial intelligence research. There, it usually refers to generating a sequence of actions that enables an \textit{agent} to achieve a specific \textit{goal}~\citep{norvig}. This can be achieved by classical search-based problem solving approaches or logical planning agents. Such planning tasks now play a significant role in a variety of demanding applications, ranging from controlling space vehicles and robots to playing the game of bridge~\citep{ghallab04}. Some of the most common planning paradigms include: (a) classical planning~\citep{wooldridge95}; (b) probabilistic planning~\citep{Bel, altman99, guo2009}; and (c) preference-based planning~\citep{son06, baier09}. Existence of a model precludes the use of each of these planning approaches. This is a limitation of all these planning approaches since not every domain has a reliable model. 

We know of at least two two kinds of planning research in software engineering. Each kind is distinguishable by {\em what} is being changed.
\bi
\item
In {\em test-based planning}, some optimization is applied to reduce the number of tests required to achieve to a certain goal or the time taken before tests yield interesting results~\citep{tallam2006concept, yoo2012regression, blue2013interaction}.
\item
In {\em process-based planning} some search-based optimizer is applied to a software process model to infer high-level business plans about software projects. Examples of that kind of work include our own prior studies sarching over  COCOMO models~\cite{me07f,Menzies:2009:ADS} or Ruhe et al.'s work on next release planning in requirements engineering~\citep{ruhe2003quantitative, ruhe2010product}. 
\ei
In software engineering, the planning problem translates to proposing changes to software artifacts. These are usually a hybrid task combining probabilistic planning and preference-based planning using search-based software engineering techniques~\citep{Harman2009, Harman2011}. These search-based techniques are evolutionary algorithms that propose actions guided by a fitness function derived from a well established domain model. Examples of algorithms used here include GALE, NSGA-II, NSGA-III, SPEA2, IBEA, MOEA/D, etc.~\citep{krall2015gale, deb00a, zit02, zit04, deb14, Cui2005a, zhang07:TEC}. 
As with traditional planning, these planning tools all require access to some trustworthy models that can be used to explore some highly novel examples. In some software engineering domains there is ready access to such models which can offer assessment of newly generated plans. Examples of such domains within software engineering include automated program repair~\citep{Weimer2009, Goues12, LeGoues2015}, software product line management~\citep{sayyad13, metzger14, henard15}, automated test generation~\citep{andrews07, andrews10}, etc. 

However, not all domains come with ready-to-use models. For example, consider all the intricate issues that may lead to defects in a product. A model that includes {\em all} those potential issues would be very large and complex. Further, the empirical data required to validate any/all parts of that model can be hard to find. Worse yet, our experience has been that accessing and/or commissioning a model can be a labor-intensive process. For example, in previous work~\citep{me07f} we used models developed by Boehm's group at the University of Southern California.Those models took as inputs project descriptors to output predictions of development effort, project risk, and defects.
Some of those models took decades to develop and mature (from 1981~\citep{boehm81} to 2000~\citep{boehm00b}). Lastly, even when there is an existing model, they can require constant maintenance lest they become out-dated. Elsewhere, we have described our extensions to the USC models to enable reasoning about agile software developments. It took many months to implement and certify those extensions~\citep{me09i, me09j}. The problem of model maintenance is another motivation to look for alternate methods that can be quickly and automatically updated whenever new data becomes available.

In summary, for domains with readily accessible models, we recommend the kinds of tools that are widely used in the search-based software engineering community such as GALE, NSGA-II, NSGA-III, SPEA2, IBEA, particle swarm optimization, MOEA/D, etc. In other cases where this is not an option, we propose the use of data mining approaches to create a quasi-model of the domain and make use of observable states from this data to generate an estimation of the model. Examples of such a data mining approaches are described below. These include five methods described in the rest of this paper: 
\bi
\item Our approaches: XTREE, BELLTREE, and 
\item Three other approaches: Alves et al.~\citep{alves}, Shatnawi~\citep{shatnawi}, and Oliveira et al.~\citep{oliveira} 
\ei

\subsection{Code based Planning}
Looking through the SE literature, we can see that researchers have proposed three methods that rely on \textit{outlier statistics} to identify suitable changes to source code metrics. The general principle underlying each of these methods is that any metric has an \textit{unusually} large (or small) value needs to be change so as not to have such large (or small) values. The key distinction between the methods is how they determine what the threshold for this unusually large  (or small) value ought to be. These methods, proposed by Alves et al.~\citep{alves}, Shatnawi~\citep{shatnawi}, and Oliveira et al.~\citep{oliveira}, are described in detail below.

\subsubsection{Alves}
Alves et al.~\citep{alves} proposed an unsupervised approach
that uses the underlying statistical 
distribution and scale of the OO metrics. It works by first weighting each metric value according to the source lines of 
code (SLOC) of the class it belongs to. All the weighted metrics are then normalized by the sum of all weights for the system. The normalized metric values are ordered in an ascending fashion (this is
equivalent a density function, where the x-axis represents 
the weight ratio (0-100\%), and the y-axis the metric scale).

Alves et al. then select a percentage value (they suggest 70\%) which 
represents the ``normal'' values for metrics. The metric threshold, then, 
is the metric value for which 70\% of the classes fall below. The 
intuition is that the worst code has outliers beyond 70\% of the normal 
code measurements i.e., they state that the risk of there existing a defect 
is moderate to high when the threshold value of 70\% is exceeded.

Here, we explore the correlation between the code metrics 
and the defect counts with a univariate logistic regression and reject 
code metrics that are poor predictors of defects (i.e.  those with $p > 
0.05$). For the remaining metrics, we obtain the threshold ranges which are denoted by $[0, 70\%)$ ranges for each metric. The plans would then involve reducing these metric range to lie within the thresholds discovered above.

\subsubsection{Shatnawi}

Shatnawi~\citep{shatnawi} offers a different alternative Alves et al by using VARL (Value of Acceptable Risk Level). This method was initially proposed by Bender~\citep{bender99} for his epidemiology studies. This approach uses two constants ($p_0$ and $p_1$) to compute the thresholds, which Shatnawi recommends to be set to $p_0=p_1=0.05$. Then using a univariate binary logistic regression three coefficients are learned:
$\alpha$ the intercept constant;
$\beta$ the coefficient for maximizing log-likelihood;
and $p_0$ to 
measure how well this model predicts for defects. (Note: the univariate 
logistic regression was conducted comparing metrics to defect counts. Any 
code metric with $p>0.05$ is ignored as being a poor defect predictor.)

Thresholds are learned from the surviving metrics using
the risk equation proposed by Bender:
$$ \mathit{Defective\ if}\ \mathit{Metric} > \mathit{VARL}$$
$$
	\mathit{VARL} = p^{-1}(p_0) = \frac{1}{\beta }\left( {\log \left( 
		{\frac{{{p_1}}}{{1 - {p_1}}}} \right) - \alpha } \right)
$$

In a similar fashion to Alves et al., we deduce the threshold ranges as $[0, VARL)$ for each selected metric. The plans would again involve reducing these metric range to lie within the thresholds discovered above.

\subsubsection{Oliveira}
Oliveira et al. in their 2014 paper offer yet another alternative to absolute threshold methods discussed above~\citep{oliveira}. Their method is still unsupervised, but they propose complementing the threshold by a second piece of information called the \textit{relative threshold}. This measure denotes the percentage of entities the upper limit should be applied to. These have the following format:
\[p\%\ of\ the\ entities\ must\ have\ M\leq k\]
Here, $M$ is an OO metric, $k$ is the upper limit of the metric value, and $p$ (expressed as \%) is the minimum percentage of entities are required to follow this upper limit. As an example Oliveira et al. state, ``85\% of the methods should have $CC \leq 14$. Essentially, this threshold expresses that high-risk methods may impact the quality of a system when they represent more than 15\% of the whole population''

The procedure attempts derive these values of $(p, k)$ for each metric $M$. They define a function \texttt{ComplianceRate(p, k)} that returns the percentage of system that follows the rule defined by the relative threshold pair $(p, k)$. They then define two penalty functions: (1) \texttt{penalty1(p, k)} that penalizes if the compliance rate is less than a constant $Min\%$, and (2) \texttt{penalty2(k)} to define the distance between $k$ and the median of preset $Tail$-th percentile. (Note: according to Oliveira et al., median of the tail is an idealized upper value for the metric, i.e., a value representing classes that, although present in most systems, have very high values of M). They then compute the total penalty as \texttt{penalty} = \texttt{penalty1(p, k)} + \texttt{penalty2(k)}. Finally, the relative threshold is identified as the pair of values $(p, k)$ that has the lowest total \texttt{penalty}. After obtaining the $(p, k)$ for each OO metric. As in the above two methods, the plan would involve ensuring the for every metric $M$ $p\%$ of the entities have a value that lies between $(0, k]$.

%% file: example.tex
% \begin{figure}[!t]
%     \centering
%     \includegraphics[width=\linewidth]{images/example.png}
%     \caption{A proposed planning framework.}
%     \label{fig:flowchart}
% \end{figure}

\begin{figure}
\centering
\captionsetup[subfigure]{width=\linewidth}
\subfloat[subfig:plans][Recommendations from some planner. 
The terms highlighted in the first row come from Figure~\ref{tab:static_metrics}.
In the second row, 
 a `$+$' represents an \textit{increase}; a `$-$'  represents an \textit{decrease};  and a `$\cdot$' represents \textit{no-change}.]{
\resizebox{0.8\linewidth}{!}{
\begin{tabular}{ccccccccc}
\hline
\rowcolor{Gray} DIT & NOC & CBO & RFC & FOUT & WMC & NOM & LOC & LCOM \\
$\cdot$ & $\cdot$    & $\cdot$    & $+$   & $\cdot$     & $+$   & $+$   & $+$   & $+$    \\\hline
\end{tabular}}
\label{subfig:plans}
}\\

\vspace{5mm}
\subfloat[][A sample of possible actions developers can take.
 Here a `$+$' represents an \textit{increase}, a `$-$' represents a \textit{decrease}, and an empty cell represents \textit{no-change}.
Taken from~\cite{stroggylos2007, du2006study, kataoka2002, bryton2009,elish2011,elish2012}. The action highlighted in \colorbox{lightgray}{gray} shows an  action matching XTREE's   recommendation
from Figure~\ref{fig:motivating_example}.A.]{
\resizebox{\linewidth}{!}{
\begin{tabular}{lccccccccc}
\hline
\rowcolor{Gray}Action                                      & DIT & NOC & CBO & RFC & FOUT & WMC & NOM & LOC & LCOM \\ 
Extract Class                               &     &     & $+$   & $-$   & $+$    & $-$   & $-$   & $-$   & $-$    \\
\rowcolor{lightgray} Extract Method                              &     &     &     & $+$   &      & $+$   & $+$   & $+$   & $+$    \\
Hide Method                                 &     &     &     &     &      &     &     &     &      \\
Inline Method                               &     &     &     & $-$   &      & $-$   & $-$   & $-$   & $-$    \\
Inline Temp                                 &     &     &     &     &      &     &     & $-$   &      \\
Remove Setting Method                       &     &     &     & $-$   &      & $-$   & $-$   & $-$   & $-$    \\
Replace Assignment                          &     &     &     &     &      &     &     & $-$   &      \\
Replace Magic Number                        &     &     &     &     &      &     &     & $+$   &      \\
Consolidate Conditional                     &     &     &     & $+$   &      & $+$   & $+$   & $-$   & $+$    \\
Reverse Conditional                         &     &     &     &     &      &     &     &     &      \\
Encapsulate Field                           &     &     &     &     &      & $+$   & $+$   & $+$   & $+$    \\
Inline Class                                &     &     & $-$   & $+$   & $-$    & $+$   & $+$   & $+$   & $+$    \\ \hline
\end{tabular}}
\label{subfig:actions}
}\\

\vspace{5mm}
\subfloat[Before `extract method'][Before `extract method']{
\fbox{\includegraphics[width=0.45\linewidth]{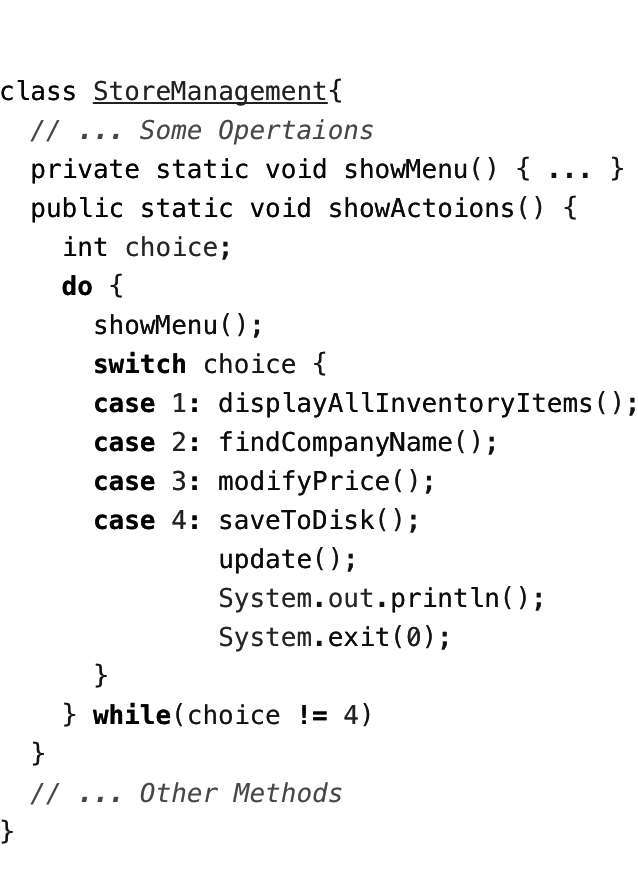}}
\label{subfig:before}
}
\subfloat[After `extract method'][After `extract method']{
\fbox{\includegraphics[width=0.45\linewidth]{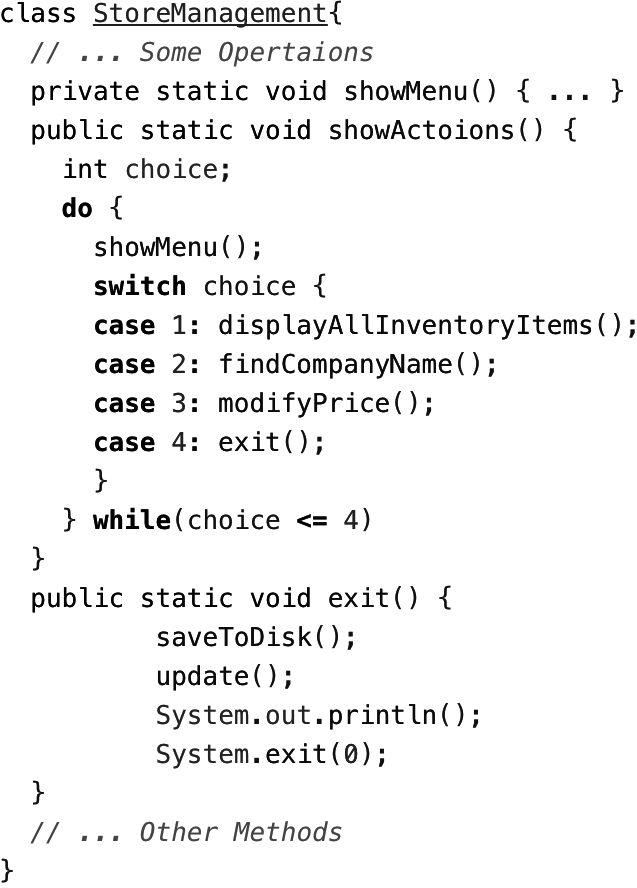}}
\label{subfig:after}
}

\caption{An example of how developers might use XTREE to reduce software defects.}
\label{fig:motivating_example}
\end{figure}

%% file: 5_xtree.tex
\section{Supervised Planning with XTREE and BELLTREE}
\input{xtree.tex}
 
 The rest of this paper comparatively evaluates:
 \bi
 \item
The value of the changes proposed
by the above methods (from Alves, Shatnawi,Oliviera et al.);
 \item
Against the changes proposed by the   XTREE/BELLTREE method described below.
\ei

\subsection{Within-Project Planning With XTREE}
\label{sect:XTREE}

Planning with XTREE is comprised of three steps namely, (a)~Frequent pattern mining; (b)~Decision tree construction; and (c)~Planning with random walk traversal. 

\noindent\textbf{Step-1: Frequent pattern mining.} The first step in XTREE is to determine which metrics are most often changed together. The OO metrics tabulated in~\tab{static_metrics} are not independent of each other. In other words, changing one metric (say $\mathit{LOC}$) would lead to a corresponding change in other metrics (such as $\mathit{CBO}$). We refrain from using correlation to determine which metrics change together because correlation measures the existence of a monotonic relationships between two metrics. We cannot assume that the metrics are monotonically related; moreover, it is possible that more than two metrics are related to each other. Therefore, we use frequent pattern mining~\citep{han2007frequent}, which represents a more generalized relationship between metrics, to detect which of the metrics change together.  

Our instrumentation is shown in~\fig{xtree}.A. We use the FP-Growth algorithm~\citep{han2007frequent} to identify the \textit{maximal frequent itemset} (highlighted in \colorbox{lightgreen}{green} in~\fig{xtree}.A-(d)). This represents the longest set of metrics that change together atleast $\mathit{support}\%$~(in our case $60\%$) of the time. The following steps use the \textit{maximal frequent itemset} to guide the generation of plans.  

\noindent\textbf{Step-2: Decision tree construction.} Having discovered which metrics change together, we next establish what range of values for each metrics point to a high likelihood of defects. For this we use a decision tree algorithm (see~\fig{xtree}.B). Specifically, we do the following: 
\be[wide=0pt]
\item Each of the OO metrics (from~\tab{static_metrics}) are discretized into a range of values with the help of Fayyad-Irani discretizer~\citep{fi}.
\item We sort the OO metrics from the most discriminative  to the least discriminative.
\item We begin by constructing a tree with the most discriminative OO metric, e.g., in~\fig{xtree}.B (b) this would be $\mathit{rfc}$.
\item Then, we repeat the above to steps on the remaing OO metrics. 
\item When we reach a predetermined termination criteria of having \textit{less than $\sqrt{N}$ samples in subsequent splits}, we do not recurse futher. Here, $N$ is the number of OO metrics, i.e., $N=20$.
\item Finally, we return the constructed decision tree.  
\ee
The leaf nodes of the decision tree contain instances of the training data that are most alike. The mean defect count of these instances represents the defect probability. In the case of~\fig{xtree}.B (b), if $\mathit{rfc}=[0,1),~\mathit{KLOC}=[3,5),~\text{and}~\mathit{DIT}=[1,6)$ then the probability of defect is $0.9$. 

\noindent\textbf{Step-3: Random Walk Traversal.} With the last two steps, we now know (1) which metrics change together and (2) what ranges of metrics indicate a high likelihood of defects. with this information, XTREE builds plans from the branches of the tree as follows. Given a ``defective'' test instance, we ask:
\be
\item
Which \underline{\textit{current}} node does the test instance fall into?
\item What are all the \underline{\textit{desired}} nodes the test case would want to emulate? These would be nodes with the \textit{lowest} defect probabilities.
\ee

Finally, we implement a random-walk~\citep{ying2018graph, sharma2016graphjet} model to find paths that lead from the \textit{current} node the \textit{desired} node. Of all the paths that lead from the \textit{current} node to the \textit{desired} node, we select the path that has the highest overlap with the \textit{maximal frequent itemset}. As an example, consider~\fig{xtree}.C. Here, of the two possible paths~\fig{xtree}.C(a) and~\fig{xtree}.C(b), we choose that latter because it traverses through all the metrics in the maximal frequent itemset.

\subsubsection*{How are plans generated?}

The path taken by the random-walk is used to generate a plan. For example, in the case of~\fig{xtree}.C, it works as follows:
\be
\item The test case finds itself on the far left, i.e., the ``current node'' has: $RFC: [0, 1)$, $KLOC: [3,5)$ and $DIT: [1,6)$
\item After implementing the random walk, we find that ``desired'' node is on the far right (highlighed in \colorbox{black}{{\color{white} black}})
\item The path taken to get from the ``current node'' to the ``desired node'' would require that the following changes be made.
\bi
\item[$\circ$] $~RFC:  [0, 1) \longrightarrow [1, 5)$;
\item[$\circ$] $KLOC:  [0, 1) \longrightarrow [1, 3)$; and
\item[$\circ$] $~CBO:  [6, 10)$
\ei
The plan would then be these ranges of values.
\ee

\subsection{Cross-project Planning with BELLTREE}
\label{sect:CPXTREE}

Many methods have been proposed for transferring data or lessons
learned from one project to another, for examples see~\citep{Nam2013, Nam2015, jing15, kocaguneli2011find, kocaguneli2012, turhan09, peters15}. Of all these, the bellwether method described here is one of the simplest.
Transfer learning with bellwethers is just a matter of calling existing
learners inside a for-loop. For all the training data from different projects $\mathcal{P, Q, R, S...}$, 
a bellwether learner conducts a round-robin experiment where a model is learned from project, then applied to all others. The {\em bellwether} is that project which generates the best performing model. The {\em bellwether effect}, states that models
learned from this bellwether performs as well as, or better than, other transfer learning algorithms. 

For the purposes of prediction, we have shown previously that bellwethers are remarkably
effective for many different kinds of SE tasks such as (i)~defect prediction, (ii)~effort
estimation, and (iii)~detecting code smells~\citep{krishna17b}. This paper is the first to check the value of bellwethers for the purposes of planning. Note also that this paper's use of bellwethers enables us to generate plans from different data sets from across different projects. This represents a novel and significant extension to our previous work~\citep{krishna17a} which was limited to the use of datasets from within a few projects.

BELLTREE extends the three bellwether operators defined in our previous work~\citep{krishna17b} on bellwethers: DISCOVER, PLAN, VALIDATE. That is:~ 
\be
  \item DISCOVER: {\em Check if a community has bellwether.} 
  This step is similar to our previous technique used to discover bellwethers~\citep{krishna16}. We see if standard data miners can predict for the number of defects, given the static code attributes. This is done as follows:~ 
  \bi 
  \item
  For a community $C$ obtain all pairs of data from
  projects $\mathcal{P, Q, R, S...}$ such that $x, y \in C$;
  \item
  Predict for defects in $y$ using a quality predictor learned from data taken from $x$;
  \item
  Report a bellwether if one $x$ generates consistently high predictions in a majority of $y \in C$.
  \ei
Note, since the above steps perform an all-pairs comparison, the theoritical complexity of the DISCOVER phase will be be $O(N^2)$ where $N$ is the number of projects.
  \item PLAN: {\em Using the bellwether, we generate plans that can improve a new project.} That is, 
  having learned the bellwether on past data, we now construct a decision tree similar to within-project XTREE. We then use the same methodology to generate the plans.
  \item VALIDATE: {\em Go back to step 1} if the performance statistics seen during PLAN fail to generate useful actions.
  \ee

%% file: xtree.tex
\begin{figure}[pt]
\centering
\arrayrulecolor{gray}
\resizebox{\linewidth}{!}{
\begin{tabular}{|p{\linewidth}|}
\hline\vspace{-0.5em}
\begin{minipage}[c]{\linewidth} 
\fig{xtree}.A: To determine which of  metrics are usually changed together, we use frequent itemset mining. Our dataset is continuous in nature (see (a)) so we first discretize  using Fayyad-Irani~\citep{fi}; this gives us a representation shown in (b). Next, we convert these into ``transactions'' where each file contains a list of discretized OO-metrics (see (c)). Then we use  the  {\em FP-growth} algorithm to mine frequent itemsets. We return the \textit{maximal frequent itemset} (as in (d)). Note: in (d) the row in \colorbox{lightgreen}{green} is the maximal frequent itemset.\\
\begin{minipage}[c]{\linewidth}
\begin{minipage}[c]{0.45\linewidth}
\resizebox{\linewidth}{!}{
\begin{tabular}{@{}c|c|c|c|c|c@{}}
    &rfc&loc&dit&cbo&Bugs  \\\hline
1.java&0.6&100&1&4&0  \\\hline
2.java&0.9&223&4&5&1  \\\hline
3.java&1.1&290&5&7&1  \\\hline
4.java&2.1&700&10&12&3  \\\hline
5.java&2.3&800&11&15&3  \\
\end{tabular}}
\end{minipage}$~~~\longrightarrow~~~$\begin{minipage}[c]{0.375\linewidth}
\resizebox{\linewidth}{!}{
\begin{tabular}{@{}c|c|c|c|c@{}}
    &rfc&loc&dit&cbo  \\\hline
1.java&A&A&A&A  \\\hline
2.java&A&A&B&A  \\\hline
3.java&A&A&B&A  \\\hline
4.java&B&B&C&B \\\hline
5.java&B&B&C&B  \\
\end{tabular}}
\end{minipage}
\end{minipage}
\begin{minipage}[c]{\linewidth}
\centering
(a)\hspace{0.5\linewidth}(b)
\end{minipage}
\begin{minipage}[c]{0.4\linewidth}
\resizebox{\linewidth}{!}{
\begin{tabular}{@{}c|c@{}}
    &Items  \\\hline
1.java&$rfc_A,~loc_A,~dit_A,~cbo_A$  \\\hline
2.java&$rfc_A,~loc_A,~dit_B,~cbo_A$  \\\hline
3.java&$rfc_A,~loc_A,~dit_B,~cbo_A$  \\\hline
4.java&$rfc_B,~loc_B,~dit_C,~cbo_B$  \\\hline
5.java&$rfc_B,~loc_B,~dit_C,~cbo_B$  \\
\end{tabular}}
\end{minipage}$~~~\longrightarrow~~~$\begin{minipage}[c]{0.45\linewidth}
\resizebox{\linewidth}{!}{
\begin{tabular}{@{}c|c@{}}
Items (\texttt{min\_sup=60})   & Support  \\\hline
$rfc_A$ & 60\\\hline
$loc_A$ & 60\\\hline
\st{$dit_A$} & \st{40}\\\hline
$\{rfc_A, loc_A\}$, $\{loc_A, cbo_A\}$, \ldots & 60 \\\hline
\rowcolor{lightgreen}$\{rfc_A, loc_A, cbo_A\}$&  60 \\\hline
\rowcolor{white}\st{$\{rfc_A, loc_A, cbo_A, dit_{B,C}\}$}&\st{40}   \\
\end{tabular}}
\end{minipage}
\begin{minipage}[c]{\linewidth}
\centering
(c)\hspace{0.5\linewidth}(d)
\end{minipage}\vspace{0.4em}
\end{minipage}\bigstrut\\\hline\vspace{-0.5em}
\begin{minipage}[c]{\linewidth}
\fig{xtree}.B: To build the decision tree, we find the most informative feature,i.e., the feature which has the lowest mean entropy of splits and construct a decision tree recursively in a top-down fashion as show below.\\
\begin{minipage}[c]{\linewidth}
\begin{minipage}[c]{0.45\textwidth}
\centering
  \begin{algorithm}[H]\scriptsize
    \caption{N-ary Decision Tree}\label{alg:dtree}
    \begin{algorithmic}[0]
    \Procedure{nary\_dtree}{train}
    \State features = train[train.columns[:-1]]
    \For{$f \in features$}
        \State Split using Fayyad-Irani method
        \State Compute entropy of splits
    \EndFor
    \State $f_{best} \gets $ Feature with least entropy
    \State Tree $\gets$ Tree.\texttt{add\_node}(f\_best)
    \State $D_{v} \gets$ Induced sub-datasets from train based on $f_{best}$
     \For{$d \in D_{v}$}
        \State Tree$_v \gets$  \textsc{nary\_dtree}(d)
        \State Tree $\gets$ Tree$_v$
    \EndFor\\
    \noindent\Return Tree
    \EndProcedure
    \end{algorithmic}
 \end{algorithm}
\end{minipage}~\begin{minipage}[c]{0.5\textwidth}
\centering
    \includegraphics[width=\linewidth]{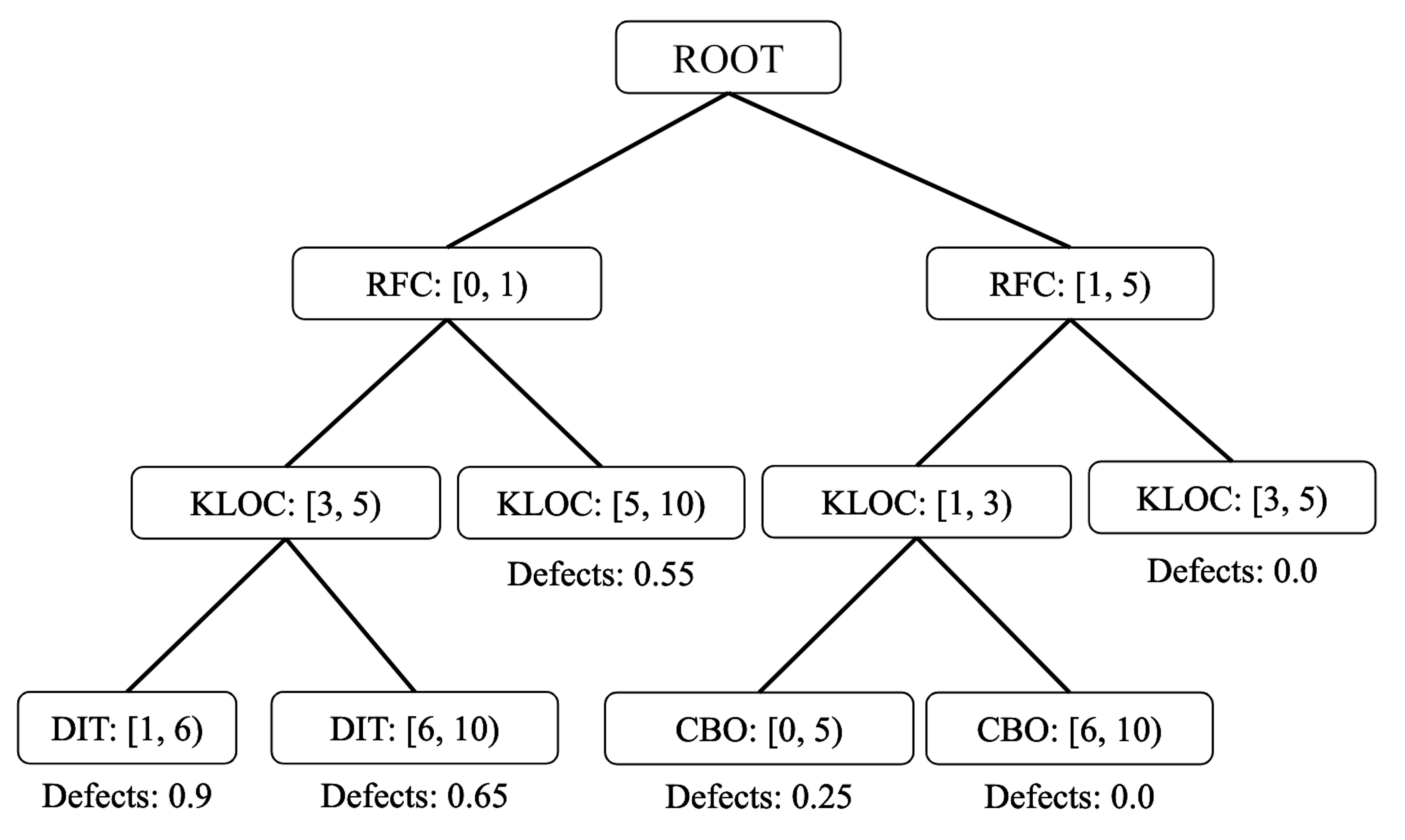}
    \label{fig:sample_figure}
\end{minipage}
\begin{minipage}[c]{\linewidth}
\centering
(a)~Decision Tree Algorithm\hspace{0.3\linewidth}(b)~Example decision tree
\end{minipage}
\end{minipage}
\end{minipage}\vspace{0.4em}
\bigstrut\\\hline\vspace{-0.5em}
\fig{xtree}.C: For ever test instance, we pass it down the decision tree constructed in \fig{xtree}.B. The node it lands is called the ``start''. Next we find all the ``end'' nodes in the tree, i.e., those which have the lowest likelihood of defects (labeled in \colorbox{black}{{\color{white} black}} below). Finally, perform a random-walk to get from ``start'' to ``end''. We use the mined itemsets from \fig{xtree}.A to guide the walk. When presented with multiple paths, we pick the one which has the largest overlap with the frequent items. e.g., in the below example, we would pick path (b) over path (a).\\[-0.25cm]
\begin{minipage}[c]{\textwidth}
\subfloat[][]{
    \includegraphics[width=0.5\linewidth]{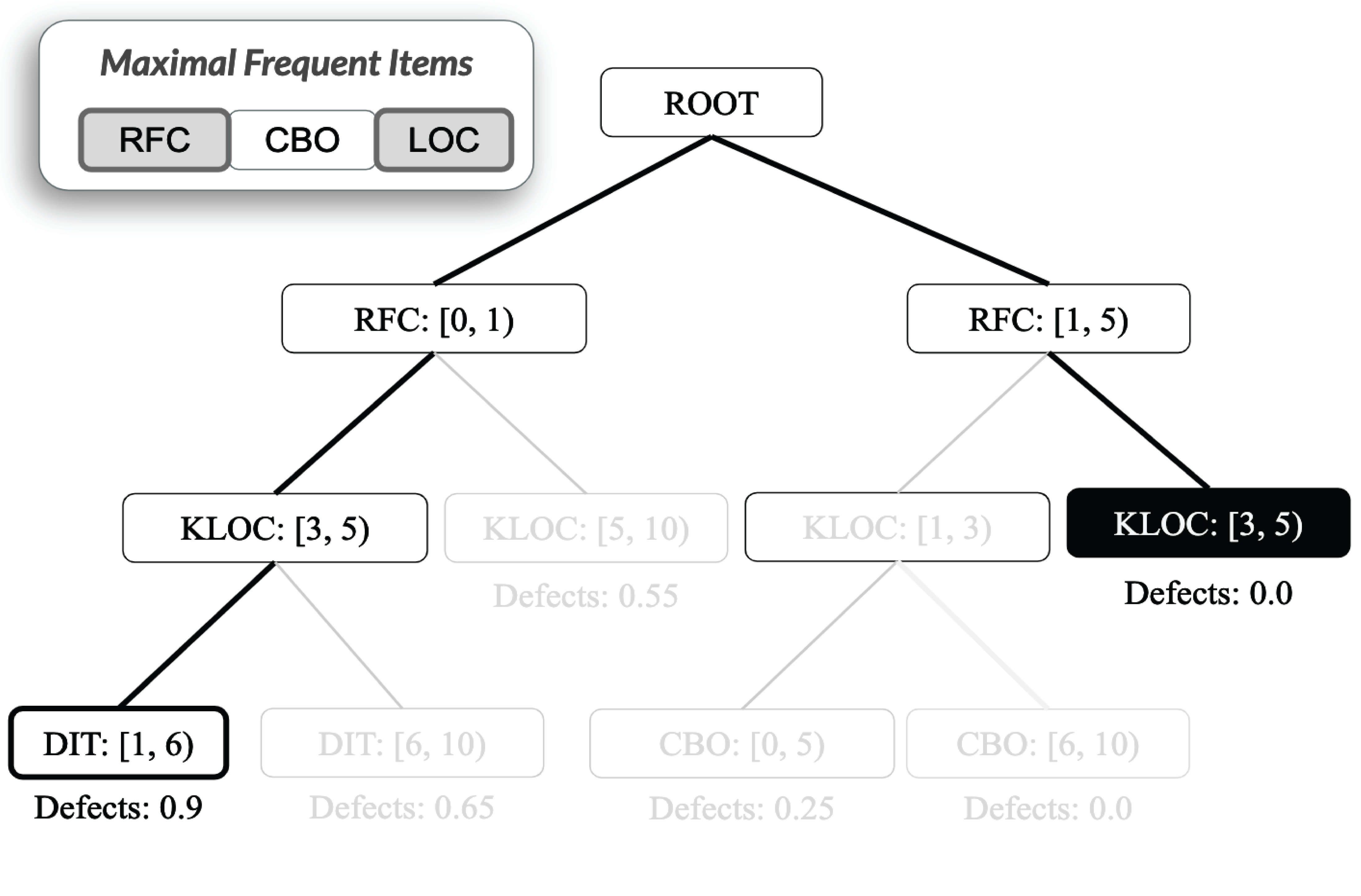}
}
\subfloat[][]{
    \includegraphics[width=0.5\linewidth]{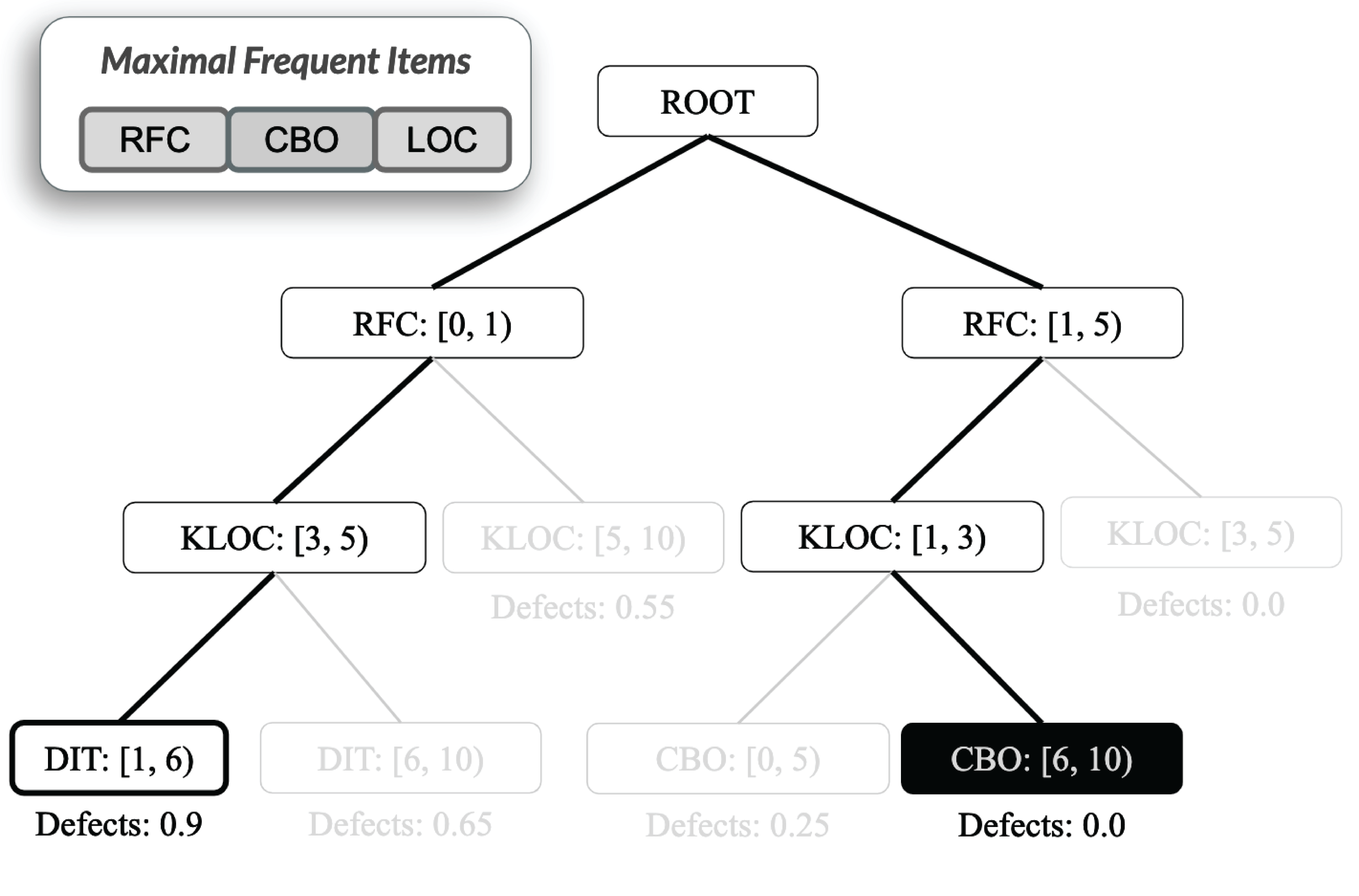}
}
\end{minipage}\bigstrut\\\hline
\end{tabular}}
\caption{XTREE Framework}
\label{fig:xtree}
\end{figure}

%% file: 6_methods.tex
\section{ Methods}
\label{sect:prelim}

The following experiment compare XTREE and BELLTREE against  
Alves, Shatnawi, Oliveira et al.

\subsection{A Strategy for Evaluating Planners}

It can be somewhat difficult to judge the effects of applying plans
to software projects. These plans cannot be assessed just by a rerun of the test suite for three reasons: (1) The defects were recorded by a post release bug tracking system. It is entirely possible it escaped detection by the existing test suite; (2) Rewriting test cases to enable coverage of all possible scenarios presents a significant challenge; and (3) It may take a significant amount of effort to write new test cases that identify these changes as they are made.

To resolve this problem, SE researchers such as
Cheng et al.~\citep{Cheng10}, O'Keefe et al.~\citep{OKeeffe08, OKeeffe07}, 
Moghadam~\citep{Moghadam2011} and Mkaouer et al.~\citep{Mkaouer14}
use a {\em verification oracle} learned separately from the primary oracle. This oracles assesses how defective the code is before and after some code changes. For their oracle, Cheng, O'Keefe, Moghadam and Mkaouer et al. use the QMOOD quality model~\citep{Bansiya02}. A shortcoming of QMOOD is that quality models learned from other projects may perform poorly when applied to new projects~\citep{localvsglobal}. As a results, we eschew using these methods in favor of evaluation strategies discussed in the rest of this section.

% As an example, consider \fig{overlap_example}; there we have 2 sets of changes: (1) Changes made by developers ($\mathcal{D}$), and (2) Changes recommended by the planner ($\mathcal{P}$). In each case we have 3 possible actions for every metric: (1) Make no change (`$\cdot$'), (2) Increase (`$+$'), and (3) Decrease (`$-$'). The intersection of the changes represents the number of times the actions taken by the developers is the same as the actions recommended by the planner. This the above example, the intersection, $\mathcal{D}\cap\mathcal{P}=7$, out of a total of $\mathcal{D}\cup\mathcal{P}=9$ possible actions. This leads to $Overlap=\frac{7}{9}\times100=77.77\%$.

\subsubsection{The \ktest}
\label{sect:ktest}

This section offers details on the evaluation method introduced
at the end of \tion{hoc}.

In order to measure the extent to which the recommendations made by planning tools matches those undertaken by the developers,
we assess the impact making those changes would have on an upcoming release of a project. For this purpose, we propose the \ktest. 

{We say that a project $\mathcal{P}$ is released in versions $\mathcal{V}\in\{\mathcal{V}_{i}, \mathcal{V}_j, \mathcal{V}_{k}\}$. Here, in terms of release dates, $\mathcal{V}_i$ precedes $\mathcal{V}_j$, which in turn precedes $\mathcal{V}_k$.} We will use these three sets for  \textit{train}, \textit{test}, and \textit{validation}, respectively\footnote{And recall in 
\tion{hoc} these versions were  given less formal names, specifically
{\em older, newer, latest}.}. These
three sets  are used as follows:

\be
\item First, train the planner on version $\mathcal{V}_{i}$. Note: this could either be data that is either from a previous release, or it could be data from the bellwether project. 

\item Next, use the planner to generate plans to reduce defects for files that were reported to be buggy in version $\mathcal{V}_{j}$.

\item Finally, on version $\mathcal{V}_{k}$, for \textit{only} the files that were reported to be buggy in the previous release, we measure the OO metrics. 
\ee

Having obtained the changes at version $\mathcal{V}_{k}$ we can now (a) measure the \textit{overlap} between plans recommended by the planner and the developer's actions, and (b) count the number of defects reduced (or possibly increased) when compared to the previous release. Using these two measures, we can assess the impact of implementing these plans. Details on measuring each of these are discussed in the subsequent parts of this section.

To compute that overlap, we proceeded as follows.
Consider  two sets of changes: 
\be
\item
$\mathcal{D}$: The changes that developers made, perhaps in response to the issues raised in a post-release issue tracking system; 
\item
$\mathcal{P}$: The plans recommended by an automated planning tool, \textit{overlap} attempts to compute the extent to which a developer's action matches that of the actions recommended by planners. 
\ee
To measure this \textit{overlap}, we use Jaccard similarity:
\begin{equation}
\setlength{\abovedisplayskip}{0pt}
\setlength{\belowdisplayskip}{0pt}
\label{eq:jaccard}
\mathit{Overlap} = \frac{|\mathcal{D} \cap \mathcal{P}|}{|\mathcal{D} \cup \mathcal{P}|}\times 100  
\end{equation}
\begin{figure}[pt!]
    \centering
    \resizebox{\linewidth}{!}{
    \begin{tabular}{c|ccccccccc}
    & DIT & NOC & CBO & RFC & FOUT & WMC & NOM & LOC & LCOM \bigstrut\\\hline
    Version $\mathcal{V}_{k}$ & 3 & 4 & 4 & 2 & 5 & 2.5 & 3 & 400 & 6 \bigstrut\\
    $\mathcal{P}\rightarrow\mathcal{V}_{k+1}$ & \cellcolor[HTML]{D0D0D0}$\cdot$ & \cellcolor[HTML]{D0D0D0}$\cdot$ & $\cdot$  & \cellcolor[HTML]{D0D0D0}$[4,7]$  & $\cdot$   & \cellcolor[HTML]{D0D0D0}$[3,6]$  & \cellcolor[HTML]{D0D0D0}$[4,7]$  & \cellcolor[HTML]{D0D0D0}$[1000, 2000]$  & \cellcolor[HTML]{D0D0D0}$[1,4]$  \bigstrut\\
    $\mathcal{D}\rightarrow\mathcal{V}_{k+1}$ & \cellcolor[HTML]{D0D0D0}$3$ & \cellcolor[HTML]{D0D0D0}$4$ & $3$  & \cellcolor[HTML]{D0D0D0}$5$  & $3$   & \cellcolor[HTML]{D0D0D0}$5$  & \cellcolor[HTML]{D0D0D0}$4$  & \cellcolor[HTML]{D0D0D0}$1500$  & \cellcolor[HTML]{D0D0D0}$2$  \\
    \end{tabular}}
    \begin{equation*}
        \setlength{\abovedisplayskip}{0em}
        \setlength{\belowdisplayskip}{-1.1em}
        Overlap = \frac{|\mathcal{D} \cap \mathcal{P}|}{|\mathcal{D} \cup \mathcal{P}|}\times 100 = \frac{7}{9}\times100 = 77.77\%
    \end{equation*}
    \caption{A simple example of computing overlap. Here a `$\cdot$' represents \textit{no-change}. Columns shaded in \colorbox{lightgray}{gray} indicate a match between developer's changes and planner's recommendations.}
    \label{fig:overlap_example}
    \end{figure}    
In other words, we measure the ratio of the size of the intersection between the developers plans and the size of all possible \textit{changes}.  Note
that  the {\em larger} the intersection between the changes made by the developers to the changes recommended by the planner, then the {\em greater} the overlap. 

An simple example of how overlap is computed is illustrated in~\fig{overlap_example}. Here, we have 9 metrics and let's say a defective file version $\mathcal{V}_{k}$ has metric values corresponding to row labeled Version $\mathcal{V}_{k}$. The row labeled $\mathcal{P}\rightarrow\mathcal{V}_{k+1}$ contains set of treatments recommended by a planner $\mathcal{P}$ for version $\mathcal{V}_{k+1}$ (note that the recommendations are ranges of values rather than actual numbers). Finally, the row labeled $\mathcal{D}\rightarrow\mathcal{V}_{k+1}$ are the result of a developer taking certain steps to possibly reduce the defects in the file for version $\mathcal{V}_{k+1}$. We see that in two cases (CBO and FOUT) the developers actions led to changes in metrics that were not prescribed by the planner. But in 7 cases, the developers actions matched the changes prescribed by the planner. Computing overlap as per~\eq{jaccard}, produces an overlap value of $77\%$.
%Note again, that we only ever measure overlap on \textit{defective} files in version $\mathcal{V}_{k}$ that were defective. There may be several files in $\mathcal{V}_{k}$ that are not defective. These non-defective files are ignored so as not bias our findings.

\subsection{Presentation of Results}

Using the {\ktest}  and overlap counts defined above,
we can measure  the overlap between the planners' recommendations and developers actions. With this, plot three kinds of charts to discuss our results:
\input{report_format.tex}
\be
\item \textit{Overlap vs. Counts}: A plot of overlap ranges (x-axis) versus the count of files that have that specific overlap range (on the y-axis). This is illustrated in \fig{sample_charts}. Here the overlap counts (x-axis) have 4 ticks: 0 (labeled 100). We see that, in the case of XTREE, the number of files that have between $76\%-100\%$ overlap is significantly larger than any other overlap range. This implies that most of the changes recommended by $XTREE$ are exactly what the developers would have actually done. On the other hand, for the other three planners (Alves, Shatnawi, and Oliveira) the number of files that have between $0\%-25\%$ overlap is significantly larger than any other overlap range. This means that those planners' recommendation are seldom what developers actually do.

\item \textit{Overlap vs. Defects reduced}: Just because there is an overlap, it does not necessarily mean that the defects were actually reduced. To measure what impact overlaps between planners' recommendations and developers actions have on reduction of defects, we plot a chart of overlap (x-axis) against the actual number of defects reduced. This is illustrated in \fig{sample_charts}. The key distinction between this chart and the previous chart is the y-axis, here the y-axis represents the number of defects reduced. Larger y-axis values for larger overlaps are desirable because this means that more the developers follow a planners' actions, higher the number of defects reduced.

\item {Overlap vs. Defects increased}: It is also possible that defects are increased as a result of overlap. To measure what impact overlaps between planners' recommendations and developers actions have on \textit{increasing} defectiveness, we plot a chart of overlap (x-axis) against the actual number of defects increased. This is illustrated in \fig{sample_charts}. The key distinction between this chart and the previous two charts is the y-axis, here the y-axis represents the number of defects \textit{increased}. Lower y-axis values for larger overlaps are desirable because this means that more the developers follow a planners' actions, lower the number of defects increased.
\ee

%% file: report_format.tex
\newcommand*\circled[1]{\tikz[baseline=(char.base)]{
            \node[circle,draw,inner sep=2pt] (char) {#1};}}
\begin{figure}
\begin{minipage}{0.49\linewidth}
\includegraphics[width=\linewidth]{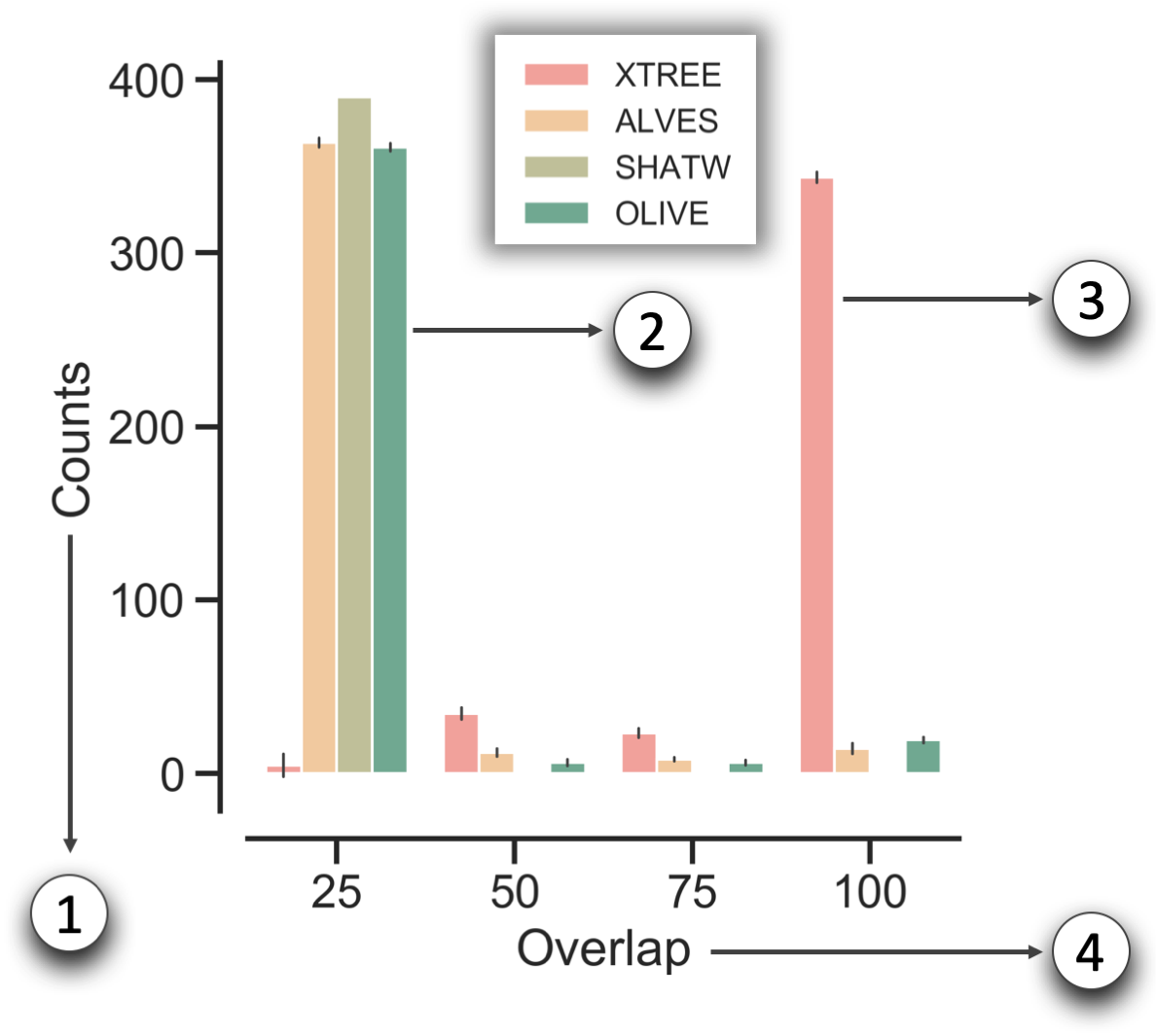}
\resizebox{1.05\linewidth}{!}{
\begin{tabular}{rl}

    \circled{1} & Counts the number of files\bigstrut[t]\\
    &\\
    \multirow{2}{*}{\circled{2}} & Number of files which have an overlap within\\
    &  the range between 0\% to 25\%\bigstrut\\
    &\\
    \multirow{2}{*}{\circled{3}} & Number of files which have an overlap within\\
    & the range between 76\% to 100\%\bigstrut\\
    &\\
    \multirow{2}{*}{\circled{4}} & Overlap ranges from 0\% to 100\% in steps\\
    &  of 25\%\bigstrut\\
    &\\
    \circled{5} & A count of \#defects removed or \#defects added.\bigstrut\\
    &\\
    \circled{6} & Number of defects removed (or added) at \bigstrut\\
    \multicolumn{1}{c}{\&}& 76\%--100\% overlap. Note: the scales are different. \bigstrut[t]\\
     \circled{7}& Usually, \#defects removed $\gg$ \#defects added.\bigstrut[b]\\
\end{tabular}}

\end{minipage}~
\begin{minipage}{0.45\linewidth}
\includegraphics[width=\linewidth]{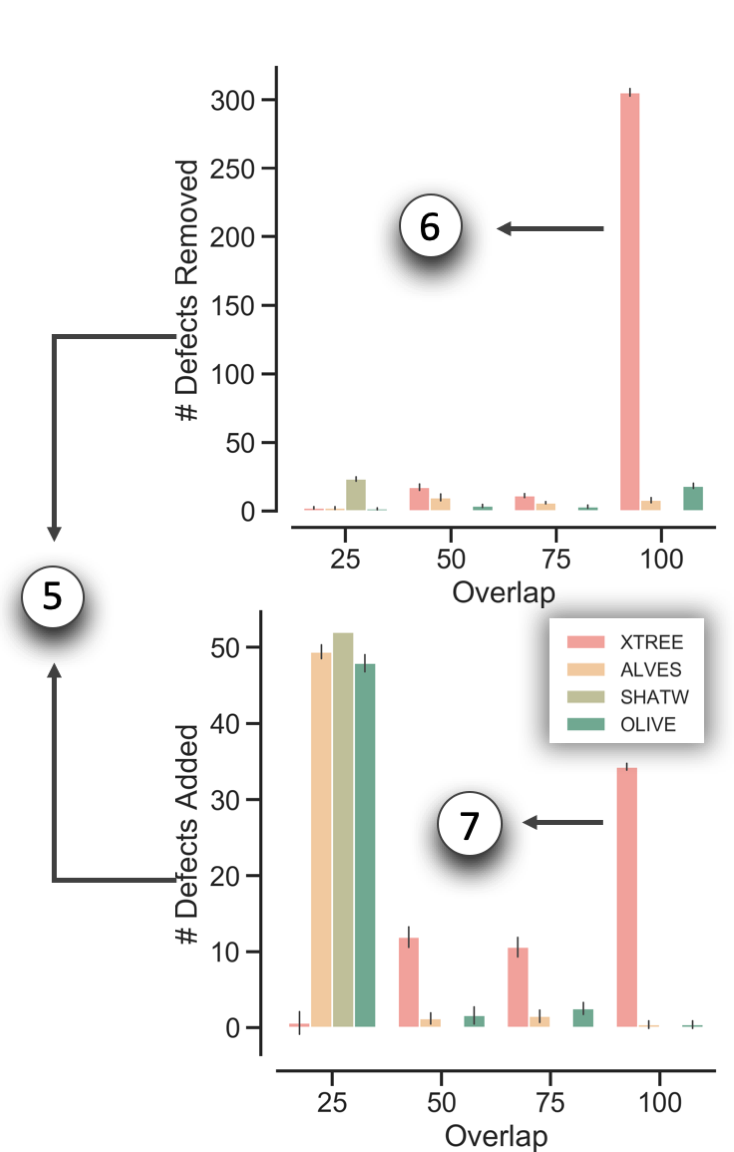}
\end{minipage}
\caption{Sample charts to illustrate the format used to present the results.}
\label{fig:sample_charts}
\end{figure}

%% file: 7_results.tex
\section{Experimental Results}
\label{sect:results}

All our experiments were conducted on a 6 core, 3.7 GHz, Intel i7-8700K running an Ubuntu 18.04 operating system. 

\subsection*{{\bf RQ1: How well do planners recommendations match  developer actions?}}
\input{RQ1.tex}

To answer this question, we measure the \textit{overlap} between the planners' recommendations and the developer's actions. To measure this, we split the available data into training, testing, and validation sets. That is, given versions $\mathcal{V}_1, \mathcal{V}_2, \mathcal{V}_3....$, we, 
\be
\item {\em train} the planners on version $\mathcal{V}_1$; then 
\item {\em generate plans} using the planners for version $\mathcal{V}_2$;
\item then {\em validate} the effectiveness of those plans on $\mathcal{V}_2$ using the \ktest.
\ee
Then,  we repeat the process by training on $\mathcal{V}_2$, testing on $\mathcal{V}_3$, and validating on version $\mathcal{V}_4$, and so on. For each of these $\{train, test, validation\}$ sets, we measure the \textit{overlap} and categorize them into 4 ranges:
\bi
\item very little, i.e. $0-25\%$;
\item some, i.e. $26\%-50\%$;
\item more, i.e. $51\%-75\%$;
\item mostly, i.e. $76\%-100\%$.
\ei
\fig{results} shows the results of planning with several planners: XTREE, Alves, Shatnawi, and Oliveira. Note, for the sake of brevity, we illustrate results for 4 projects-- Ant, Camel, Poi, and Xerces. A full set set results for all projects are available at  \url{https://git.io/fjkNM}. 

We observe a clear dichotomy in our results. 
\bi
\item All outlier statistics based planners (i.e., those of Alves, Shatnawi, and Oliveira) have overlaps only in the range of 0\% to 25\%. This means that \textit{most of the developers actions did not match the recommendations proposed by these planners.}
\item In the case of XTREE, the largest number of files had an overlap of 76\% to 100\% and second largest was between 51\% to 75\%. This means that, in a majority of cases developers actions are 76\% to 100\% similar to XTREE's recommendations. At the very least, there was an 51\% similarity between XTREE's recommendations and developers actions.
\ei
We observe this trend in all 18 datasets-- XTREE significantly outperformed other threshold based planners in terms of the overlap between the plans and the actual actions undertaken by the developers. Note that reason the results are very negative about the methods of Alves, Shatnawi, Oliveira, et al. is because their recommendations would be very hard to operationalize (since those recommendations were seldom seen in the prior history of a project). Thus, our response to this research question can be summarized as follows:
\input{RQ2.tex}

\result{XTREE significantly outperforms all the other outlier statistics based planners. Further, in all the projects studied here, most of the developer actions to fix defects in a file has as 76\%--100\% overlap with the recommendations offered by XTREE.}

\subsection*{{\bf RQ2: Do planners' recommendation lead to reduction in defects?}}

In the previous research question measured the extent to which a planner's recommendations matched the actions taken by developers to fix defects in their files. But, the existence of a high overlap in most files does not necessarily mean that the defects are actually reduced. Likewise, it is also conceivable that that defects are added due to other actions the developer took during their development. Thus, it is important to ask how many defects are reduced, and how many are added, in response to larger overlap with the planners' recommendations.

Our experimental methodology to answer this research question is as follows: 
\bi
\item Like before, we measure the \textit{overlap} between the planners' recommendations developers' actions. 
\item Next, we plot the aggregate number defects reduced and in file with overlap values ranging from 0\% to 100\% in bins of size 25\% (for ranges of $0-25\%$, $26-50\%$, $51-75\%$, and $76-100\%$). 
\ei

Note that, we refrain from computing the correlation between $\mathit{overlap}$ and defects increased/decreased because we were interested only in the cases with large overlaps, i.e., cases where overlap $>75\%$. In these cases, we measure what impact the changes have on bug count. Correlation, we disovered, was ill-suited for this purpose because it does not distinguish between low-/high-overlaps it only measures the linearity of the relationship between overlap and defect count. For example, in an ideal case where every plan offered by XTREE is followed, the overlaps at 0\% --- 99\% would be zero and so would the value of correlation, but would be most misleading. 

Similar to RQ1, we compare XTREE with three other outlier statistics based planners of Alves et al., Shatnawi, and Oliveira, for the overall number of defects reduced and number of defects added. We prefer planners that have a large number defects reduced for higher overlap ranges are considered better.

\fig{rq2} shows the results of planning with several planners: XTREE, Alves, Shatnawi, and Oliveira. Note that, similar to the previous research question, we only illustrate results for 4 projects-- Ant, Camel, Poi, and Xerces. A full set of results for RQ2 for all projects are available at \url{https://git.io/fjIvG}. 

We make the following observations from in our results: 
\be
\item \textit{Defects Decreased}: \fig{rq2}\protect\subref{fig:rq2_a} plots the number of defects \textit{removed} in files with various overlap ranges. It is desirable to see larger defects removed with larger overlap. We note that:
\bi
\item When compared to other planners, the number of defects removed as a result of recommendations obtained by XTREE is significantly larger. This trend was noted in all the projects we studied here.
\item In the cases of Ant, Camel, and Xerces there are large number of defect reduced when the overlap lies between 76\% and 100\%. Poi is an exception-- here, we note that the largest number of defects are removed when the overlap is between 51\% and 75\%. 
\ei

\item \textit{Defects Increased}: \fig{rq2}\protect\subref{fig:rq2_b} plots the number of defects \textit{added} in files with various overlap ranges. It is desirable to see lower number of defects added with larger overlap. We note that: 
\bi
\item When compared to other planners, the number of defects added as a result of recommendations obtained by XTREE is comparatively larger. This trend was noted in all the projects we studied here. This is to be expected since, developers actions seldom match the recommendations of these other planners. 

\item In all the cases the number of defects removed was significantly larger than the number of defects added. For example, in the case of Camel, 420+ defects were removed at 76\% -- 100\% overlap and about 70 defects were added (i.e., 6$\times$ more defects were removed than added). Likewise, in the case of Xerces, over 300 defects were removed and only about 30 defects were added (i.e., 10$\times$ more defects were removed than added).
\ei
\ee

The ratio of defects removed to the number of defects added is very important to asses. \fig{rq2_1} plots this ratio at 76\% -- 100\% overlap (it applied equally for the other overlap ranges as they have far fewer defects removed and added). From this chart, we note that out of 18 datasets, in 14 cases XTREE lead to a significant reduction in defects. For example, in the case of Ivy and Log4j, there were no defects added at all.

However, in 4 cases, there were more defects added than there were removed. Given the idiosyncrasies  of real world projects, we do not presume that developers will always take actions as suggested by a planner. This may lead to defects being increased, however, based on our results we notice that this is not a common occurrence.
In summary, our response to this research question is as follows:

\noindent\result{Plans  generated  by  XTREE  are  superior  to  other  outlier  statistics based  planners  in  all  10  projects.  Planning  with  XTREE  leads  to  the  far larger number of defects reduced as opposed to defects added in 9 out of 10 projects studied here.}
\input{RQ2_1.tex}
\vspace{-0.4em}

\subsection*{{\bf RQ3: Are  cross-project  plans  generated  by  BELLTREE  as  effective  as  within-project plans of XTREE?}}
\input{RQ3.tex}

In the previous two research questions, we made an assumption that there are past releases that can be used to construct the planners. However, this may not always be the case. For new project, it is quite possible that there are not any historical data to construct the planners. In such cases, SE literature proposes the use of \textit{transfer learning}. In this paper, we leverage the so-called \textit{bellwether} effect to identify a bellwether project. Having done so, we construct a planner quite similar to XTREE with the exception that the training data comes from the bellwether project. This variant of our planner that uses the bellwether project is called the BELLTREE (see \tion{CPXTREE} for more details).

To answer this research question, we train XTREE on within-project data and generate plans for reducing the number of defects. We then compare this with plans derived from the bellwether data and BELLTREE. We hypothesized that since bellwethers have been demonstrated to be efficient in prediction tasks, learning from the bellwethers for a specific community of projects would produce performance scores comparable to within-project data. Our experimental methodology to answer this research question is as follows: 
\be
\item Like before, we measure the \textit{overlap} between the planners' recommendations developers' actions. 
\item Next, we tabulate the aggregate number defects reduced (\fig{rq3}\protect\subref{fig:rq3_dec}) and the number of defects increased (\fig{rq3}\protect\subref{fig:rq3_inc}) in files with overlap values ranging from 0\% to 100\% in bins of size 25\% (for ranges of $0-25\%$, $26-50\%$, $51-75\%$, and $76-100\%$). 
\ee

Similar to previous research questions, we compare XTREE with BELLTREE and a random oracle (RAND). We prefer planners that have a large number defects reduced for higher overlap ranges and planner that have lower number of defects added are are considered better.

We make the following observations from in our results: 
\be
\item \textit{Defects Decreased}: \fig{rq2}\protect\subref{fig:rq2_a} plots the number of defects \textit{removed} in files with various overlap ranges. It is desirable to see larger defects removed with larger overlap. We note that:
\bi
\item When compared to other planners, the number of defects removed as a result of recommendations obtained by XTREE is significantly larger. This trend was noted in all the projects we studied here.
\item In the cases of Ant, Camel, and Xerces there are large number of defect reduced when the overlap lies between 76\% and 100\%. Poi is an exception-- here, we note that the largest number of defects are removed when the overlap is between 51\% and 75\%. 
\ei

\item \textit{Defects Increased}: \fig{rq2}\protect\subref{fig:rq2_b} plots the number of defects \textit{added} in files with various overlap ranges. It is desirable to see lower number of defects added with larger overlap. We note that: 
\bi
\item When compared to other planners, the number of defects added as a result of recommendations obtained by XTREE is comparatively larger. This trend was noted in all the projects we studied here. This is to be expected since, developers actions seldom match the recommendations of these other planners. 

\item In all the cases the number of defects removed was significantly larger than the number of defects added. For example, in the case of Camel, 420+ defects were removed at 76\% -- 100\% overlap and about 70 defects were added (i.e., 6$\times$ more defects were removed than added). Likewise, in the case of Xerces, over 300 defects were removed and only about 30 defects were added (i.e., 10$\times$ more defects were removed than added).
\ei
\ee

The ratio of defects removed to the number of defects added is very important to asses. \fig{rq2_1} plots this ratio at 76\% -- 100\% overlap (it applied equally for the other overlap ranges as they have far fewer defects removed and added). From this chart, we note that out of 18 datasets, in 14 cases XTREE lead to a significant reduction in defects. For example, in the case of Ivy and Log4j, there were no defects added at all.

However, in 4 cases, there were more defects added than there were removed. Given the idiosyncrasies  of real world projects, we do not presume that developers will always take actions as suggested by a planner. This may lead to defects being increased, however, based on our results we notice that this is not a common occurrence.

In summary, our response to this research question is as follows:\\

\result{The effectiveness of BELLTREE and XTREE are similar. If within-project data is available, we recommend using XTREE. If not, BELLTREE is a viable alternative.}

% \subsection*{{\bf RQ4: How many changes do the planners propose?}}
% \input{deltas.tex}

% This question naturally follows the findings of the previous research questions. Here, we ask how many changes each of the planners recommend. This is important because having plans recommend far too many changes would make it challenging for practical use. 

% Our findings for XTREE tabulated in \fig{deltas}\footnote{Space limitations prohibit showing results of BELLTREE. We notice a very similar trend to XTREE. Interested readers can use our replication package (\url{https://git.io/fNcYY}) to further evaluate these results.} show that XTREE (BELLTREE) proposes far fewer changes compared to other planners. This is because, both XTREE and BELLTREE operate based on supervised learning incorporating two stages of data filtering and reasoning: (1) Discretization of attributes based on information gain, and (2) Plan generation based on contrast sets between adjacent branches. This is different to the other approaches. The operating principle of the other approaches is that attribute values larger than a certain threshold must always be reduced. Hence, they usually propose plans that use all attributes in an unsupervised manner, without first filtering out the less important attributes based on how they impact the quality of software. This leads to those planners being far more verbose and, possibly, harder to operationalize.

% \result{Our planning methods (XTREE/BELLTREE) recommend far fewer changes than other methods.}

%% file: RQ1.tex
\def\rot{\rotatebox}
\newcolumntype{a}{>{\columncolor{gr1}}r}
\newcolumntype{b}{>{\columncolor{gr2}}r}
\newcolumntype{c}{>{\columncolor{gr3}}r}
\newcolumntype{d}{>{\columncolor{gr4}}r}
\begin{figure}[!htbp]
\centering
\includegraphics[width=0.75\linewidth]{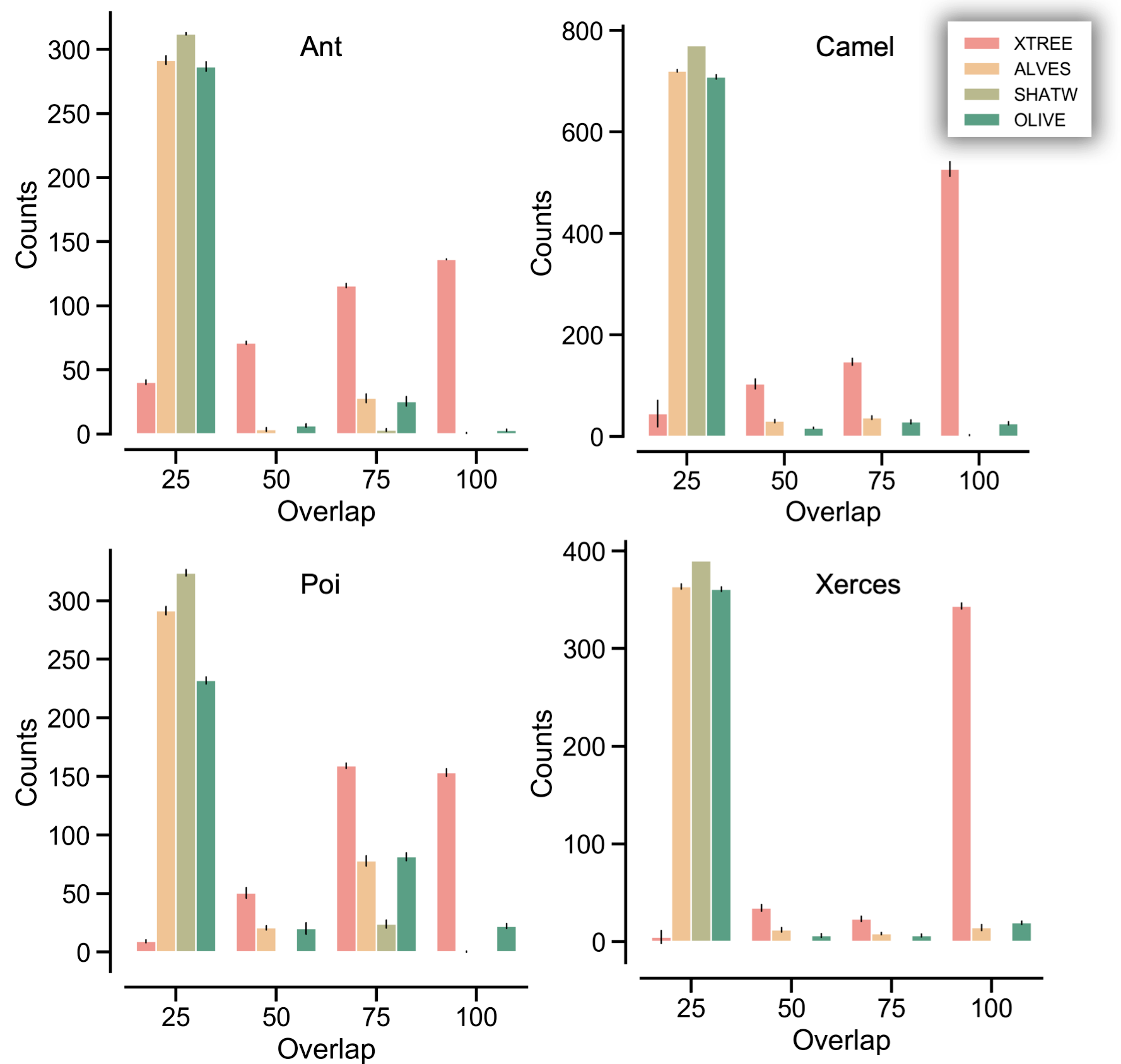}
\caption{A count of number of test instances where the developer changes overlaps a planner recommendation. The overlaps (in the x-axis) are categorized into four ranges for every dataset (these are $0\leq~Overlap\leq25$, $26\leq~Overlap\leq50$, $51\leq~Overlap\leq75$, and $76\leq~Overlap\leq100$). For each of the overlap ranges, we count the the number of instances in the validation set where overlap between the planner's recommendation and the developers changes fell in that range. Note: \textit{Higher counts} for larger overlap is \textit{better}, e.g.,  $Count([75,100]) > Count([0,25))$ is considered better.}
\label{fig:results}
\end{figure}

%% file: RQ2.tex
\def\rot{\rotatebox}
\newcolumntype{a}{>{\columncolor{gr1}}r}
\newcolumntype{b}{>{\columncolor{gr2}}r}
\newcolumntype{c}{>{\columncolor{gr3}}r}
\newcolumntype{d}{>{\columncolor{gr4}}r}
\begin{figure}[!htbp]
\centering 
\subfloat[Defects Reduced]{
\includegraphics[width=0.75\linewidth]{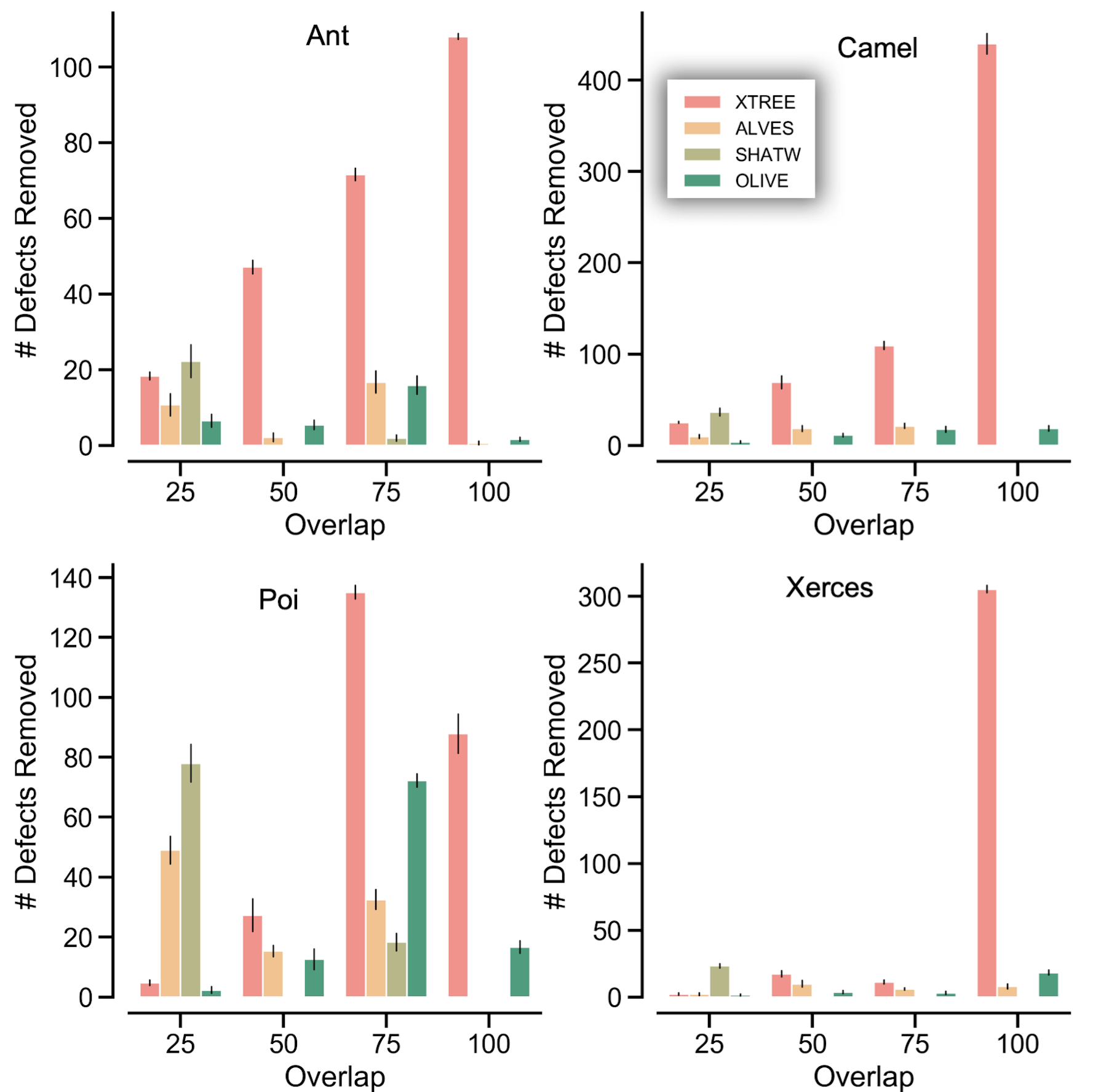}
\label{fig:rq2_a}}\\~\hrule~
\subfloat[Defects Increased]{
\includegraphics[width=0.75\linewidth]{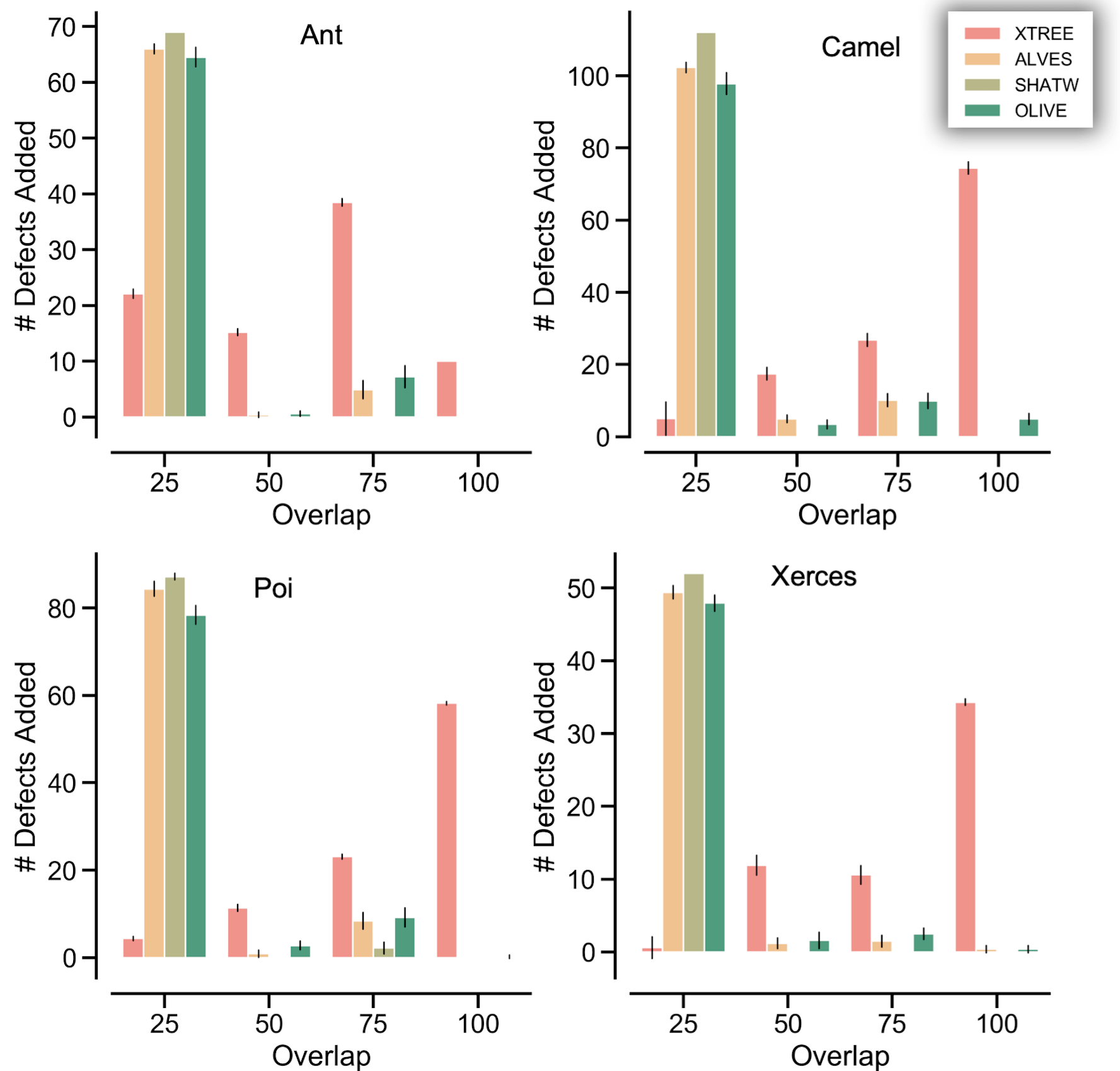}
\label{fig:rq2_b}}
\caption{A count of total number \textit{defects reduced} and \textit{defects increased} as a result each planners' recommendations. The overlaps are again categorized into four ranges for every dataset (denoted by $min\leq~Overlap<max$). For each of the overlap ranges, we count the total number of \textit{defects reduced} and \textit{defects increased} in the validation set for the classes that were defective in the test set as a result of overlap between the planner's recommendation and the developers changes that fell in the given range}
\label{fig:rq2}
\end{figure}

%% file: RQ2_1.tex
\begin{figure}[!tbp]
\centering 
\includegraphics[width=0.75\linewidth]{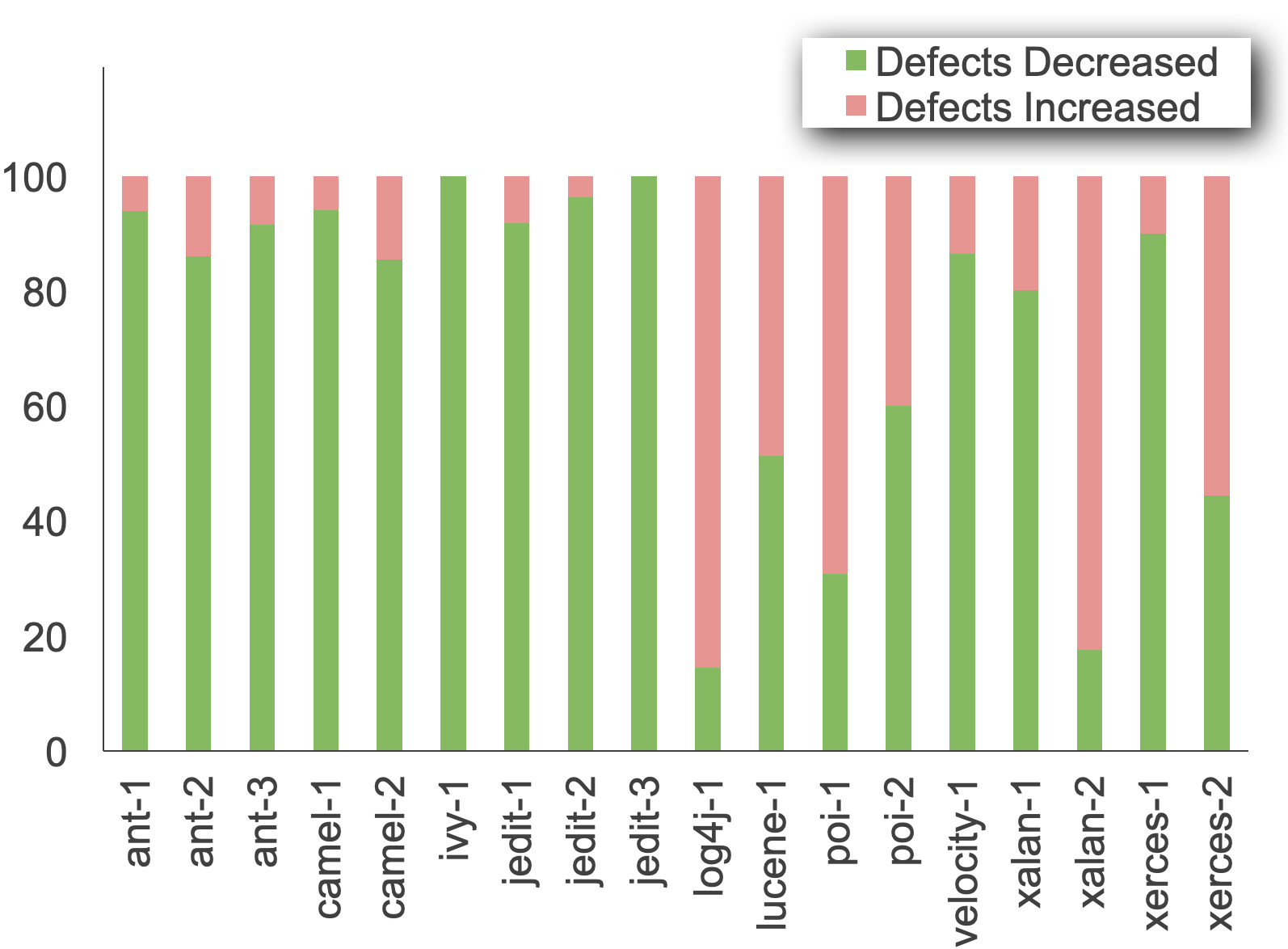}
\caption{A count of total number \textit{defects reduced} and \textit{defects increased} as a result each planners' recommendations. The overlaps are again categorized into four ranges for every dataset (denoted by $min\leq~Overlap<max$). For each of the overlap ranges, we count the total number of \textit{defects reduced} and \textit{defects increased} in the validation set for the classes that were defective in the test set as a result of overlap between the planner's recommendation and the developers changes that fell in the given range}
\label{fig:rq2_1}
\end{figure}

%% file: RQ3.tex
\def\rot{\rotatebox}
\begin{figure}[!htbp]
\centering 
\subfloat[subfig:plans][Defects Reduced. \textit{Higher defect reduction} for larger Overlap is considered \textit{better}.]{
\arrayrulecolor[gray]{0.5}
\resizebox{0.8\linewidth}{!}{
\begin{tabular}{@{}l|rrr|rrr|rrr|rrr}
            \rowcolor{white}& \multicolumn{3}{c|}{$[0, 25)$~~~~~~}       & \multicolumn{3}{c|}{$[25, 50)$~~~~~}       & \multicolumn{3}{c|}{$[50, 75)~~~~~$}       & \multicolumn{3}{c}{$[75, 100]$~~~~}       \bigstrut\\ \hline
            &  \rot{90}{RAND} & \rot{90}{XTREE} & \rot{90}{BELLTREE~~} & \rot{90}{RAND} & \rot{90}{XTREE} & \rot{90}{BELLTREE~~} & \rot{90}{RAND} & \rot{90}{XTREE} & \rot{90}{BELLTREE~~} & \rot{90}{RAND} & \rot{90}{XTREE} & \rot{90}{BELLTREE~~}  \bigstrut\\\hline
ant-1      & \cellcolor{gr1}13    & \cellcolor{gr1}13    & 12       & 3     & \cellcolor{gr2}33    & 19       & 0     & \cellcolor{gr4}27    & 10       & 0     & \cellcolor{gr4}62    & 54       \bigstrut\\

ant-2      & 0     & 6     & \cellcolor{gr1}13       & 0     & \cellcolor{gr2}42    & 33       & 0     & \cellcolor{gr4}27    & \cellcolor{gr4}27       & 0     & \cellcolor{gr4}124   & 61       \bigstrut\\

ant-3      & \cellcolor{gr1}22    & 18    & 6        & 1     & \cellcolor{gr2}71    & 42       & 0     & \cellcolor{gr4}47    & 27       & 0     & 108   & \cellcolor{gr4}124      \bigstrut\\ \hline

camel-1    & \cellcolor{gr1}76    & 29    & 10       & 0     & \cellcolor{gr2}90    & 30       & 0     & \cellcolor{gr3}52    & 20       & 0     & \cellcolor{gr4}226   & 98       \bigstrut\\

camel-2    & \cellcolor{gr1}36    & 25    & 30       & 0     & \cellcolor{gr2}109   & 100      & 0     & \cellcolor{gr3}69    & 68       & 0     & \cellcolor{gr4}439   & 277      \bigstrut\\ \hline

ivy-1      & 1     & 4     & \cellcolor{gr1}12       & 0     & 10    & \cellcolor{gr2}42       & 0     & 5     & \cellcolor{gr3}13       & 0     & 12    & \cellcolor{gr4}25       \bigstrut\\ \hline

jedit-1    & \cellcolor{gr1}13    & 9     & 11       & 8     & 35    & \cellcolor{gr2}44       & 0     & 39    & \cellcolor{gr3}50       & 0     & \cellcolor{gr4}136   & 108      \bigstrut\\

jedit-2    & \cellcolor{gr1}28    & 24    & 10       & 1     & \cellcolor{gr2}77    & 34       & 0     & 36    & \cellcolor{gr3}39       & 0     & 107   &\cellcolor{gr4} 135      \bigstrut\\

jedit-3    & 18    & \cellcolor{gr1}30    & 28       & 1     & 67    & \cellcolor{gr2}75       & 0     & 28    & \cellcolor{gr3}35       & 0     & 70    & \cellcolor{gr4}106      \bigstrut\\ \hline

log4j-1    & \cellcolor{gr1}5     & 1     & 0        & 0     & 7     & \cellcolor{gr2}14       & 0     & 3     & \cellcolor{gr3}8        & 0     & 8     & \cellcolor{gr4}50       \bigstrut\\ \hline

poi-1      & 1     & 0     & \cellcolor{gr1}7        & 5     & 0     & \cellcolor{gr2}80       & 0     & 2     & \cellcolor{gr3}19       & 0     & 81    & \cellcolor{gr4}90       \bigstrut\\

poi-2      & \cellcolor{gr1}78    & 4     & 0        & 18    & \cellcolor{gr2}135   & 0        & 0     & \cellcolor{gr3}27    & 2        & 0     & \cellcolor{gr4}87    & 83       \bigstrut\\ \hline

velocity-1 & \cellcolor{gr1}51    & 2     & 6        & 0     & \cellcolor{gr2}25    & 15       & 0     & \cellcolor{gr3}39    & 32       & 0     & \cellcolor{gr4}90    & 48       \bigstrut\\ \hline

xalan-1    & \cellcolor{gr1}22    & 6     & 2        & \cellcolor{gr2}105   & 43    & 51       & 13    & 60    & \cellcolor{gr3}66       & 0     & \cellcolor{gr4}409   & 230      \bigstrut\\

xalan-2    & \cellcolor{gr1}110   & 0     & 6        & 0     & 38    & \cellcolor{gr2}49       & 0     & \cellcolor{gr3}102   & 54       & 0     & 83    & \cellcolor{gr4}408      \bigstrut\\ \hline

xerces-1   & \cellcolor{gr1}23    & 2     & 11       & 0     & 11    & \cellcolor{gr2}13       & 0     & 17    & \cellcolor{gr3}24       & 0     & \cellcolor{gr4}305   & 49       \bigstrut\\

xerces-2   & \cellcolor{gr1}7     & 0     & 2        & 0     & 3     & \cellcolor{gr2}11       & 0     & 6     & \cellcolor{gr3}18       & 0     & 117   & \cellcolor{gr4}305     

\end{tabular}}
\label{fig:rq3_dec}}\\
\subfloat[subfig:plans][Defects Increased. In comparison to defects reduced in \fig{rq2}(a) above, we would like to have as little defects increased as possible.]{
\arrayrulecolor[gray]{0.5}
\resizebox{0.8\linewidth}{!}{
\begin{tabular}{@{}l|rrr|rrr|rrr|rrr}
            \rowcolor{white}& \multicolumn{3}{c|}{$[0, 25)$~~~~~~}       & \multicolumn{3}{c|}{$[25, 50)$~~~~~}       & \multicolumn{3}{c|}{$[50, 75)$~~~~~}       & \multicolumn{3}{c}{$[75, 100]$~~~~}       \bigstrut\\ \hline
            &  \rot{90}{RAND} & \rot{90}{XTREE} & \rot{90}{BELLTREE~~} & \rot{90}{RAND} & \rot{90}{XTREE} & \rot{90}{BELLTREE~~} & \rot{90}{RAND} & \rot{90}{XTREE} & \rot{90}{BELLTREE~~} & \rot{90}{RAND} & \rot{90}{XTREE} & \rot{90}{BELLTREE~~}  \bigstrut\\ \hline
ant-1      & 15  & \cellcolor{gr1}1  & \cellcolor{gr1}3 & 0  & \cellcolor{gr2}10 & \cellcolor{gr2}11 & 0 & \cellcolor{gr3}2  & \cellcolor{gr3}1  & 0 & 4   & \cellcolor{gr4}2   \bigstrut\\
ant-2      & 63  & 9  & \cellcolor{gr1}1 & 0  & 33 & \cellcolor{gr2}10 & 0 & 11 & \cellcolor{gr3}2  & 0 & 20  & \cellcolor{gr4}4   \bigstrut\\
ant-3      & 69  & 22 & \cellcolor{gr1}9 & 0  & 38 & \cellcolor{gr2}33 & 0 & 15 & 11 & 0 & \cellcolor{gr4}10  & 20  \bigstrut\\ \hline
camel-1    & 36  & 10 & \cellcolor{gr1}5 & 0  & \cellcolor{gr2}11 & 25 & 0 & \cellcolor{gr3}6  & 14 & 0 & \cellcolor{gr4}14  & 31  \bigstrut\\
camel-2    & 112 & \cellcolor{gr1}5  & \cellcolor{gr1}2 & 0  & 26 & \cellcolor{gr2}15 & 0 & 17 & \cellcolor{gr3}\cellcolor{gr3}9  & 0 & 74  & \cellcolor{gr4}15  \bigstrut\\ \hline
ivy-1      & 6   & \cellcolor{gr1}1  & \cellcolor{gr1}0 & 0  & \cellcolor{gr2}3  & \cellcolor{gr2}2  & 0 & \cellcolor{gr3}2  & \cellcolor{gr3}1  & 0 & 0   & 0   \bigstrut\\ \hline
jedit-1    & 37  & \cellcolor{gr1}3  & \cellcolor{gr1}2 & 2  & 20 & \cellcolor{gr2}10 & 0 & 11 & \cellcolor{gr3}6  & 0 & 12  & \cellcolor{gr4}6   \bigstrut\\
jedit-2    & 15  & \cellcolor{gr1}2  & \cellcolor{gr1}5 & 0  & \cellcolor{gr2}8  & 19 & 0 & \cellcolor{gr3}2  & 11 & 0 & \cellcolor{gr4}4   & 12  \bigstrut\\
jedit-3    & 3   & \cellcolor{gr1}1  & \cellcolor{gr1}1 & 0  & 1  & \cellcolor{gr2}0  & 0 & \cellcolor{gr3}1  & \cellcolor{gr3}1  & 0 & 0   & 0   \bigstrut\\ \hline
log4j-1    & 73  & \cellcolor{gr1}1  & \cellcolor{gr1}2 & 1  & 14 & \cellcolor{gr2}7  & 0 & 13 & \cellcolor{gr3}2  & 0 & 47  & \cellcolor{gr4}7   \bigstrut\\ \hline
poi-1      & 190 & \cellcolor{gr1}1  & \cellcolor{gr1}1 & 6  & 7  & \cellcolor{gr2}1  & 0 & 5  & \cellcolor{gr3}0  & 0 & 182 & \cellcolor{gr4}6   \bigstrut\\
poi-2      & 87  & 4  & \cellcolor{gr1}0 & 2  & 23 & \cellcolor{gr2}7  & 0 & 11 & \cellcolor{gr3}5  & 0 & \cellcolor{gr4}58  & 184 \bigstrut\\ \hline
velocity-1 & 21  & \cellcolor{gr1}1  & \cellcolor{gr1}4 & 4  & \cellcolor{gr2}3  & 14 & 0 & \cellcolor{gr3}3  & 17 & 0 & 14  & \cellcolor{gr4}10  \bigstrut\\ \hline
xalan-1    & 152 & \cellcolor{gr1}2  & \cellcolor{gr1}3 & 21 & 46 & \cellcolor{gr2}29 & 6 & 33 & \cellcolor{gr3}31 & 0 & \cellcolor{gr4}101 & 217 \bigstrut\\
xalan-2    & 506 & 27 & \cellcolor{gr1}3 & 0  & \cellcolor{gr2}25 & 48 & 0 & 87 & \cellcolor{gr3}32 & 0 & 388 & \cellcolor{gr4}101 \bigstrut\\ \hline
xerces-1   & 52  & \cellcolor{gr1}0  & \cellcolor{gr1}0 & 0  & 10 & \cellcolor{gr2}1  & 0 & 11 & \cellcolor{gr3}1  & 0 & \cellcolor{gr4}34  & 1   \bigstrut\\
xerces-2   & 169 & 4  & \cellcolor{gr1}0 & 0  & 14 & \cellcolor{gr2}11 & 0 & \cellcolor{gr3}9  & 12 & 0 & 146 & \cellcolor{gr4}34 
\end{tabular}}
\label{fig:rq3_inc}
}
\caption{A count of total number \textit{defects reduced} and \textit{defects increased} as a result each planners' recommendations. The overlaps are again categorized into four ranges for every dataset (denoted by $min\leq~Overlap<max$). For each of the Overlap ranges, we count the total number of \textit{defects reduced} and \textit{defects increased} in the validation set for the classes that were defective in the test set as a result of Overlap between the planner's recommendation and the developers changes that fell in the given range}
\label{fig:rq3}
\end{figure}

%% file: 8_discuss.tex
\section{Discussion}
\label{sect:discuss}
When discussing these results with colleagues, we are often asked the following questions.

\textit{1. Why use automatic methods to find quality plans? Why not just use domain knowledge; e.g. human expert intuition?} Recent research has documented the wide variety of conflicting opinions among software developers, even those working within the same project. According to Passos et al.~\citep{passos11}, developers often assume that the lessons they learn from a few past projects are general to all their future projects. They comment, ``past experiences were taken into account without
much consideration for their context''. Jorgensen and Gruschke~\citep{jorgensen09} offer a similar warning. They report that the supposed software engineering ``gurus'' rarely use lessons from past projects to improve their future reasoning and that such poor past advice can be detrimental to new projects~\citep{jorgensen09}. Other studies have shown some widely-held views are now questionable given new evidence. Devanbu et al. examined responses from 564 Microsoft software developers from around the world. They comment programmer beliefs can vary with each project, but do not necessarily correspond with actual evidence in that project~\citep{prem16}. Given the diversity of opinions seen among humans, it seems wise to explore automatic oracles for planning.

\textit{2. Does using BELLTREE guarantee that software managers will never have to change their plans?} No. Software managers should evolve their policies when the evolving circumstances require such an update. But how to know when to retain current policies or when to switch to new ones? Bellwether method can answer this question.

Specifically, we advocate continually retesting the bellwether's status against other data sets within the community. If a new bellwether is found, then it is time for the community to accept very different policies. Otherwise, it is valid for managers to ignore most the new data arriving into that community.

%% file: 9_threats.tex
\section{Threats to Validity}
\label{sect:threats}
\begin{itemize}[leftmargin=-1pt]
\item[] \textit{Sampling Bias}: Sampling bias threatens any classification experiment;
what matters in one case may or may not hold in another case. 
For example, data sets in this study come from several sources, but they were all supplied by individuals. Thus, we have documented our selection procedure for data and suggest that researchers
try a broader range of data.
\item[] \textit{Evaluation Bias}:
This paper uses one measure for the quality of the planners
and other quality measures may be used to quantify the effectiveness of planner. A comprehensive analysis using these measures may be performed with our replication package. Additionally, other measures can easily be added to extend this replication package.

\item[] \textit{Order Bias}: 
Theoretically, with prediction tasks involving learners such as random forests, there is invariably some degree of randomness that is introduced by the algorithm. To mitigate these biases, researchers, including ourselves in our other work, report the central tendency and variations over those runs with some statistical test. However, in this case, all our approaches are \textit{deterministic}. Hence, there is no need to repeat the experiments or run statistical tests. Thus, we conclude that while order bias is theoretically a problem, 
it is not a major problem in the particular case of this study.
\end{itemize}

%% file: 10_conclusion.tex
\section{Conclusions and Future Work}
\label{sect:future}
% It is quite evident that there is a rapid growth of the use of data analytics in software engineering. Such a growth has revealed some open issues that need to be tackled. This paper is an attempt to address these issues. 

Most software analytic tools that are currently in use today are mostly prediction algorithms. These algorithms are limited to making predictions. We extend this by offering ``planning'': a novel technology for prescriptive software analytics. Our planner offers users a guidance on what action to take in order to improve the quality of a software project. Our preferred planning tool is BELLTREE, which performs cross-project planning with encouraging results. With our BELLTREE planner, we show that it is possible to reduce several hundred defects in software projects. 

It is also worth noting that BELLTREE is a novel extension of our prior work on (1) the bellwether effect, and (2) within-project planning with XTREE. In this work, we show that it is possible to use bellwether effect and within-project planning (with XTREE) to perform cross-project planning using BELLTREE, without the need for more complex transfer learners. Our results from~\fig{results} show that BELLTREE is just as good as XTREE, and both XTREE/BELLTREE are much better than other planners. 

Hence our overall conclusion is to endorse the use of planners like XTREE (if local data is available) or BELLTREE (otherwise).

% Finally, we note that BELLTREE can offer stable solutions. As long as the bellwether data from which BELLTREE is constructed remains unchanged, so would plans that are derived from it. In this sense, practitioners can expect stable plans for relatively longer periods of time. This is in contrast to our XTREE planner. As more within-project data is gathered,  it is important that XTREE planner be updated. Such constant updates would invariably lead to unstable and often contradicting plans.

% As for future work, we would like to undertake the following tasks:
% \be
% \item \textit{Industrial validation}: One of immediate goals is to validate the usefulness of these planners in a realistic development environment. As the first steps towards this, we are currently collaborating with a software company in RTP to deploy XTREE and BELLTREE planners in their pipeline. 
% % \item \textit{Developer survey: }It would also be valuable to solicit developers' feedback on planners such as XTREE/BELLTREE. A study of their willingness to use such tools would greatly benefit the software analytics community.
% \item \textit{Scaling planners:} It must be noted that the datasets studied here are relatively small. In order to be able to use XTREE/BELLTREE as a real time planner for very large projects. It is very important to scale these planners to accommodate very large datasets.
% \ee